\documentclass[reprint,showkeys,showpacs,preprintnumbers,nofootinbib,amsmath,amssymb,aps,prd,longbibliography,onecolumn,10pt]{revtex4-1}
\usepackage{array,multirow,graphicx}
\usepackage{dcolumn}
\usepackage{txfonts}
\usepackage{newlfont}
\usepackage{times}
\usepackage{bm}
\usepackage[colorlinks,citecolor=blue,urlcolor=blue,linkcolor=blue]{hyperref}
\usepackage[figtopcap]{subfigure}
\usepackage{color}

\begin{document}
\title{Charged spherically symmetric Taub-NUT black hole solutions in  $f(R)$ gravity}
\author{G.G.L. Nashed$^{1,2}$}%
\email{nashed@bue.edu.eg}
\author{Kazuharu Bamba$^{3}$}%
\email{bamba@sss.fukushima-u.ac.jp}
\affiliation{$^{1}$Centre for Theoretical Physics, The British University in Egypt, P.O. Box 43, El Sherouk City, Cairo 11837, Egypt}
\affiliation{$^{2}$Egyptian Relativity Group (ERG), Cairo University, Giza 12613, Egypt}
\affiliation{$^{3}$Division of Human Support System, Faculty of Symbiotic Systems Science, Fukushima University, Fukushima 960-1296, Japan}

\begin{abstract}
$f(R)$ theory is a modification of Einstein general relativity which has many interesting results in cosmology and astrophysics. To derive  black hole solution in this theory is difficult due to the fact that it is fourth order differential equations.  In this study, we use the first reliable  deviation from general relativity which is given by the quadratic form of $f(R)=R+\beta R^2$, where $\beta$ is a dimensional parameter.  We calculate the energy conditions of the charged black holes and show that  all of them are satisfied for the Taub-NUT spacetime. Finally, we study some thermodynamic quantities such as entropy, temperature, specific heat and Gibbs free energy.  The calculations of heat capacity and free energy show that the charged Taub-NUT black hole have positive values  which means that it has thermal stability.

\end{abstract}
\maketitle
\section{\bf Introduction}

The Hilbert-Einstein Lagrangian which involves higher order corrections of Ricci scalar has been used a long time ago as a result of the quantum correction to the gravitational field of the matter source \cite{Birrell:1982ix}.  Due to this feature, it was thought that such terms are successful to describe the early epoch of the growth of our universe. The successful model which describes this period is the Starobinsky one which explains the inflation epoch successfully \cite{Starobinsky:1980te}.  Psaltis et al. \cite{Psaltis:2007cw} have obtained  important results to test general relativity (GR)  in the
strong-field regime using astrophysical black holes.

Recently, observations confirm that our universe suffers from acceleration. Since that, the correspondence between the accelerated epoch and the inflationary mechanism help scientists to assume that dark energy may have a geometric origin \cite{Capozziello:2003tk,Carroll:2003wy,Nojiri:2003ft}. Later, scientists discovered that the terms which are related to the quantum higher order corrections are responsible for the accelerating expansion rate of our universe at large structures. This phenomenon has been investigated at the fourth order of $f(R)$ \cite{DeFelice:2010aj,2010RvMP...82..451S,Vignolo:2018eco,Capozziello:2007ec,Nojiri:2010wj,Nojiri:2017ncd,Capozziello:2011et,Capozziello:2010zz,Bamba:2015uma} (for more references on modified gravity theories as well as dark energy problem, see, e.g., \cite{Cai:2015emx,Bamba:2012cp}).  In this study, we are going to consider $f(R) = R + \beta R^2$ which we   consider as a physical model due to the following: For this theory, in the early universe, the $R^2$ term is dominant and it can lead to the so-called Starobinsky inflation (the $R^2$ inflation) \cite{Starobinsky:1980te} (for review on inflation in modified gravity theories, see, e.g., \cite{Bamba:2015uma}). It is known that  Starobinsky inflation can be consistent with the recent Planck result \cite{Akrami:2018odb}.
Moreover, in the late-time universe, the cosmological constant becomes dominant and can play a role of dark energy, so that the accelerated expansion of the universe can occur at the present time \cite{Amendola:2006eh}. Thus, in the model of $f(R) = R + \beta R^2$, both inflation in the early universe and the late-time cosmic acceleration can be realized in a unified manner. This is why we consider the $R^2$ model including a cosmological constant in the Lagrangian.

 The mechanism which determines the difference between $f(R)$ gravitational theories and the Einstein  GR, that is ensured to be detected at the scales of astrophysics and strong gravitational field, is to derive  black holes that are different from those of GR (vacuum or electro-vacuum) \cite{PhysRevD.92.083014,Bambi:2013nla,PhysRevD.83.104002,Awad:2017tyz,2018CQGra..35b5018C,Will:2004xi}.  An identification to  modified gravitation theories that can intrinsically share the same solutions with
GR has been investigated in \cite{Motohashi:2018wdq}. There are many applications carried out on $f(R)$, among them are:  The one-loop effective action of $f(R)$ theories on the de-Sitter background that has been explained in \cite{Cognola:2005de}. One of the most interesting phenomena which can occur in charged black hole solutions is the anti-evaporation process which explains the primordial black hole. In order that the anti-evaporation may occur in the Einstein theory, one needs to involve the quantum correction terms from the matter field. Anti-evaporation could occur at the classical level on modified gravity theories like $f(R)$ gravity \cite{Nojiri:2013su,Nojiri:2014jqa}. A subject to acquire solutions for static spherically symmetric black holes in $f(R)$ gravity has been addressed in \cite{PhysRevD.80.124011}. In $f(R)$ gravity, solutions for spherically symmetric black holes have been obtained by assuming that the scalar curvature is constant \cite{PhysRevD.86.024013,2013PhyS...87d5004S,2013CQGra..30l5003N}. Also, spherically symmetric non-charged and charged black hole solutions have been discussed with no constraints on the scalar curvature and the Ricci tensor in \cite{Nashed:2018piz,2018EPJP..133...18N,2018IJMPD..2750074N}. One of the merits of $f(R)$ theory is the fact that it is able to investigate the epoch from inflation to the accelerated expansion  \cite{Nojiri:2010wj}. Moreover, $f(R)$ theory has been tested using many cosmological and astrophysical applications  \cite{DeFelice:2010aj,Capozziello:2011et,Nashed:2019tuk} as well as many local tests to constrain it \cite{Sotiriou:2005xe,PhysRevD.74.023529,Nojiri:2006gh,Sawicki:2007tf}. It has been shown that the successful modified gravitational theory is the one that could describe the evolution of our universe, from the big bang to the present time,  and to be consistent with the astrophysical prediction given by GR \cite{PhysRevD.74.087501,PhysRevD.82.023519}.

In spite of the fact that  the gravitational field equations of $f(R)$ gravity are of the fourth order and therefore the non-linear terms become too complicated, many great successful efforts have been achieved. Here, we give a brief summary of these successful efforts:
Through the method with the Lagrange multiplier, in $f(R)$ gravity, a Lagrangian for the gravitational field equations has been derived in spherically symmetric spacetimes \cite{Sebastiani:2010kv}. The Friedmann universe filled with a perfect fluid has been investigated for $f(R) = R^{1+\gamma}$ with $\gamma$ a constant \cite{PhysRevD.72.103005}. This study has demonstrated that the derivatives in the gravitational field equations are at most the first order and it has obtained novel analytic solutions \cite{Sebastiani:2010kv}. In addition, static interior solutions for spherically symmetric spacetime have been found in \cite{Shojai:2011yq}. With the Noether symmetries, the solutions for static spherically symmetric black holes have been analyzed \cite{Capozziello2012}. It has been pointed out that if the universe is filled with a barotropic fluid, the expression of $f(R)$ cannot be determined by the time evolution of the scale factor \cite{PhysRevD.74.064022}. With the Weyl's canonical coordinates, the static solutions for axially symmetric black holes in vacuum have been acquired \cite{ GutierrezPineres:2012yd}. Furthermore, $d$-dimensional static spherically symmetric black holes have been obtained using the generator method \cite{Amirabi:2015aya}. For more references on the static spherically symmetric black holes, we refer to \cite{PhysRevD.90.084011,Hendi:2011eg,Nashed:2005kn,PhysRevD.84.084006,Cembranos:2011sr,Resco:2016upv,Nashed:2006yw,PhysRevD.80.104012,Azadi:2008qu,Nashed:2009hn,Nashed:2008ys,Capozziello:2007id,Nashed:2007cu,Nashed:2015pda,PhysRevD.91.104004,Nojiri:2013su,Hendi:2013zba,Hendi:2014mba,Nashed:uja,Nashed:2015qza,PhysRevD.81.124051,Hendi:2012nj,Myrzakulov:2015kda,PhysRevD.92.124019,PhysRevLett.114.171601,PhysRevD.75.027502} and references therein. Also, Psaltis et al.  and Motohashi and Minamitsuji \cite{Psaltis:2007cw, Motohashi:2018wdq} who  are succeeded to put constraints on the modified gravitational theories  so the one can easily derive the solutions of GR. However, until now, in $f(R)$ gravity, there is no study on the solutions for Taub-NUT black hole. Therefore, the main purpose of this work is to use the non-charged and charged gravitational field equations for $f(R)$ gravity theory with its quadratic form in Taub-NUT spacetime and try to find  new solutions.

The properties of gravitomagnetic  of the Taub-NUT spacetime are characterized by the non-diagonal  metric whose source comes from
 the NUT parameter $\chi$. This non-diagonal expression yields a singularity
on the half-axis $\theta=\pi$, which is called Misner string, that is different from  ordinary coordinate singularity which is related to the use
of spherical coordinates. Misner \cite{1963JMP.....4..924M} supposed to elude  such singularity is to include a periodic time
coordinate and two coordinate patches. The first one covers the northern hemisphere and the singularity
is located along the axis $\theta=\pi$, while the other patch covers the southern hemisphere with the singularity extending
along the axis $\theta=0$.   The price  paid
for a Taub-NUT spacetime to be free from these axial singularities is the use of a periodic time coordinate \cite{Kagramanova_2010}.

The use of  periodic identification of the time coordinate makes the Taub-NUT solution has a problem for physical
applications because of causality violations.  Bonnor \cite{1969PCPS...66..145B}  suggested another explanation of the Taub-NUT space-time,
to avoid the un-physical property of the Misner  explanation. He preserved the singularity at $\theta=\pi$, and endowed it
 with a semi-infinite massless rotating rod. In this investigation, the NUT space-time is  being created by
a spherically symmetric body of mass $M$ at the origin and by a source of pure angular momentum which is uniformly distributed along the $\theta=\pi$–axis.

The organization of the present work is as the following. In Section \ref{S2}, the ABC of $f(R)$ gravity is presented. The gravitational field equations for $f(R)$ gravity with its quadratic form are applied to the Taub-NUT spacetime and derived their non-linear differential equations. We solve this system of differential equations and derive an exact solution. In Section \ref{S3}, the charged gravitational field equations in $f(R)$ gravity is given and applied to the Taub-NUT spacetime considered in Section \ref{S2}. We also solve the charged field equations analytically and derive exact solutions.
In Section \ref{S4}, the relevant physics of black holes derived in Section \ref{S2} and \ref{S3} are analyzed by calculating their singularities and energy conditions. In Section \ref{S5}, we investigate thermodynamics  of black holes derived in Sections \ref{S2} and \ref{S3} and demonstrate that locally, the solution of charged black holes is stable. In the final section, we discuss the results of the present study.

\section{ABC of $f(R)$ gravity} \label{S2}
The Lagrangian of $f(R)$ gravity has the form
\begin{equation} \label{lag}
{\cal L}_g:=\frac{1}{2\kappa} \int d^4x \sqrt{-g} (f(R)-\Lambda),\end{equation}
where $\kappa =8G\pi$ is the Einstein gravitational constant and $G$ is the Newtonian gravitational constant. In Eq. (\ref{lag}), $R$ is the Ricci scalar, $g$ is the determinant for the metric tensor $g_{\mu \nu}$ and $f(R)$ is a function which is analytic and differentiable. The variations of the Lagrangian (\ref{lag}) in terms of $g_{\mu \nu}$ lead to the gravitational field equations in vacuum \cite{Cognola:2005de,Koivisto:2005yc}
\begin{equation} \label{fe11}
S_{\mu \nu}=R_{\mu \nu} f_R-\frac{1}{2}g_{\mu \nu}f(R)-2g_{\mu \nu}\Lambda +g_{\mu \nu} \Box f_R-\nabla_\mu \nabla_\nu f_R\equiv0,\end{equation}
where  $R_{\mu \nu}$ is the Ricci tensor.
 The  operator $\Box$ is the D'Alembertian operator that is defined as $\Box= \nabla_\alpha\nabla^\alpha $ where $\nabla_\alpha W^\beta$ is the covariant derivatives of the vector $W^\beta$ and $f_R=\frac{df(R)}{dR}$.  The trace of  equation (\ref{fe11}) leads to:
\begin{equation} \label{tras1}
Rf_R-2f(R)-8\Lambda+3\Box f_R=0.\end{equation} The solution of Eq. (\ref{tras1})  for constant Ricci scalar has the form \cite{Jaime:2010kn,Canate:2015dda}  \begin{equation} \label{tras}
R=-8\Lambda.
\end{equation}
 Finally, since the
power-law of $f(R)$  is the one with best agreement with   cosmological data
  \cite{Nunes:2016drj,Jana:2018djs}, therefore, in the following sections we are going to focus our
attention  to the choice
\begin{equation}\label{powellaw}
 {f(R)=R+\beta R^2},
\end{equation}
{ with $\beta$  being the dimensional model parameter}.

\subsection{Taub-NUT spacetime}
We consider spacetimes whose metric can be written locally in the form
\begin{equation} \label{m2222}
ds^2= -s(r)dt^2+\frac{1}{N_1(r)}dr^2+k(r)d\Sigma^2+2\chi s(r)k_1\left(\theta\right) [dt-2\chi k_1\left(\theta\right) d\phi]d\phi,
\end{equation}
where $s(r)$, $k(r)$ and $N_1(r)$ are arbitrary functions and  $\chi$ is  the Taub-NUT parameter. Here  $d\Sigma^2$ is a 2-dimensional Einstein-K\"ahler manifold, which can be taken to be the unit
sphere $S^2$, torus $T^2$ or the hyperboloid $H^2$ which  respectively have the form \cite{Mann:2005mb}:
\begin{eqnarray}
k_1(\theta)&=& \Biggl\{
\begin{array}{cl}
\cos\theta , & ~~~~\mathrm{{for~~ \delta=1~~sphere},\nonumber} \\
\theta , & ~~~~\mathrm{{for ~~\delta=0~~ torus},\nonumber} \\
\cosh\theta, & ~~~~\mathrm{for~~ \delta=-1~~hyperboloid}.%
\end{array}%
\end{eqnarray}
Here in this study we are interested in the case of unit sphere  $S^2$. Therefore the above metric takes the form
\begin{equation} \label{m22}
ds^2= -s(r)dt^2+\frac{1}{N_1(r)}dr^2+k(r)\left(d\theta^2+\sin^2\theta d\phi^2\right)+2\chi s(r)\cos\theta [dt-2\chi \cos\theta d\phi]d\phi,
\end{equation}
which is  the Taub-NUT spacetime.
By substituting Eq. (\ref{fe11}) into Eq. (\ref{m22}), we acquire
\begin{eqnarray} \label{fe3}
  & &S_t{}^t=\frac{\aleph(k^2N_1s'^2
-k^2ss'N'_1-2k^2sN_1s''-2ksN_1k's' -8s^3\chi^2+8k^2s^2\Lambda)}{4k^2s^2}=0,\nonumber\\
& &S_r{}^r=\frac{\aleph(k^2N_1s'^2-2k^2sN_1s''-sk^2s'N'_1+2s^2N_1k'^2-4ks^2N_1k''-2ks^2k'N'_1+8k^2s^2\Lambda)}{4s^2k^2}
=0,\nonumber\\
 && S_\theta{}^\theta=S_\phi{}^\phi= \frac{\aleph(2skN_1k''+kk'(sN_1)'-8s^2\chi^2-4ks[1+2k\Lambda])}{4sk^2}=0,\nonumber\\
& &S_\phi{}^t= \frac{\chi\aleph\cos\theta(2ks^2N_1k''-2k^2sN_1s''+k^2N_1s'^2-skk'(s'N_1-sN'_1)-k^2ss'N'_1-4s^2(2\chi^2s+k))}{2k^2s^2}=0,
\end{eqnarray}
where $\aleph=(1-16\beta \Lambda)$ and we have used Eq. (\ref{tras}).
The solution for Eq. (\ref{fe3}) when $\aleph\neq 0$ become
 \begin{eqnarray} \label{s3}
& &k(r)=\varrho^2,\qquad \qquad s(r)=s_1(r)N_1(r), \qquad \textrm {where} \qquad  s_1(r)=\frac{r^2}{c_1\varrho^2-\chi^2}, \nonumber\\
 & &N_1(r)=\frac{\{3c_2c_1{}^3\sqrt{c_1\varrho^2-\chi^2}+c_1{}^2\varrho^2[2\Lambda\varrho^2+3]+2c_1\chi^2[4\Lambda\varrho^2-3]-16\chi^4\Lambda\}
 [c_1\varrho^2-\chi^2]}{3c_1{}^3r^2\varrho^2}\, ,\nonumber\\
 & & \end{eqnarray}
 where $$\varrho=\sqrt{r^2+\chi^2}.$$
 It is of interest to note that when the constant $c_1=1$  the function $s_1(r)=1$ and the functions $s(r)$ and $N_1(r)$ will be identical.    The horizon of Eq. (\ref{s3}) is shown in figure \ref{Fig:1}\subref{fig:1a}, for $r>0$, that corresponds to \textbf{a black hole horizon}.
 The metric  of solution   (\ref{s3}) takes the form
\begin{eqnarray} \label{metr3}
ds^2&=&-\left\{\frac{3c_2c_1{}^3\sqrt{c_1\varrho^2-\chi^2}+c_1{}^2\varrho^2[2\Lambda\varrho^2+3]+2c_1\chi^2[4\Lambda\varrho^2-3]-16\chi^4\Lambda
 }{3c_1{}^3\varrho^2}\right\}dt^2+\varrho^2d\theta^2
\nonumber\\
& &+\frac{3c_1{}^3r^2\varrho^2 }{(c_1\varrho^2-\chi^2)[3c_2c_1{}^3\sqrt{c_1\varrho^2-\chi^2}+c_1{}^2\varrho^2[2\Lambda\varrho^2+3]+2c_1\chi^2[4\Lambda\varrho^2-3]-16\chi^4\Lambda]}dr^2\nonumber\\
& &-\left\{\frac{3c_1{}^3\varrho^4 \sin^2\theta-12\chi^2 c_2c_1{}^3\cos^2\theta\sqrt{c_1\varrho^2-\chi^2}-4\chi^2\cos^2\theta(c_1{}^2\varrho^2[2\Lambda\varrho^2+3]-2c_1\chi^2[4\Lambda\varrho^2-3]
-16\chi^4\Lambda)}{3c_1{}^3\varrho^2}\right\}d\phi^2\nonumber\\
& &+4\chi\cos\theta\left\{\frac{3c_2c_1{}^3\sqrt{c_1\varrho^2-\chi^2}+c_1{}^2\varrho^2[2\Lambda\varrho^2+3]
+2c_1\chi^2[2\Lambda\varrho^2-3]-16\chi^4\Lambda}{3c_1{}^3\varrho^2}\right\}dtd\phi\;.
\end{eqnarray}
 Equation (\ref{metr3}) shows that solution (\ref{s3}) behaves asymptotically as AdS/dS. When $c_1=1$, the metric spacetime  (\ref{metr3}) is singular at $\varrho=0$ and  $r=0$. ~Moreover,  the metric (\ref{metr3}) has another singularity at \begin{eqnarray} \label{scal33} 3c_2r+\varrho^2[2\Lambda\varrho^2+3]+2\chi^2[4\Lambda\varrho^2-3]-16\chi^4\Lambda=0.\end{eqnarray}  Equation (\ref{scal33}) is a fourth algebraic equation that has two real positive solutions.

 Calculating all the invariants of solution (\ref{s3}) we get
\begin{eqnarray} \label{scal3} && R^{\mu \nu \lambda \rho}R_{\mu \nu \lambda \rho}=\frac{\mathbb{F}_1(r)}{3\varrho^{12}},\qquad \qquad \qquad R^{\mu \nu}R_{\mu \nu}=16\Lambda^2,
\end{eqnarray}
 with $\mathbb{F}_1(r)$ being a lengthy polynomial function and we have put $c_1=1$ to make the line element (\ref{metr3}) has a well-know asymptote behavior.  Equation (\ref{scal3}) shows that a true singularity exists when $\varrho=0\Rightarrow r=0$.   It is interesting  to note that the singularity that arise from Eq. (\ref{scal33} ) and makes the metric (\ref{metr3}) singular does not make the  Kretschmann invariant and squared Ricci of Eq. (\ref{scal3}) divergent.

\section{Charged Taub-NUT black hole solution}\label{S3}
The Lagrangian of $f(R)$ gravity with a coupling between geometry and matter takes the form
\begin{equation} \label{lag1}
{\cal L}:={\cal L}_g+{\cal L}_{em},
\end{equation}
with ${\cal L}_g$ being the Lagrangian of the gravitation that is given by Eq. (\ref{lag}). The Lagrangian of matter is given by
\begin{equation}  \label{lag1}
{{\cal L}}_{
em}:=-\frac{1}{2}{ F}\wedge {^{\star}F},
\end{equation}
where $F = dA$ with $A=A_{\mu}dx^\mu$ the $1$-form of the gauge potential \cite{Awad:2017tyz} and ${^{\star}F}$ is the dual of $F$. The variations of the action (\ref{lag1}) in terms of $g_{\mu \nu}$ and the vector potential $A_\mu$ give the field equation in the Maxwell-$f(R)$ gravity as \cite{Cognola:2005de,Koivisto:2005yc}:
\begin{equation}  \label{fec}
R_{\mu \nu} f_R-\frac{1}{2}g_{\mu \nu}f(R)-2g_{\mu \nu}\Lambda +g_{\mu \nu} \Box f_R-\nabla_\mu \nabla_\nu f_R=2\kappa T_{\mu \nu}, \qquad \qquad
\partial_\nu \left( \sqrt{-g} {\textsl F}^{{^{\mu \nu}}} \right)=0,
\end{equation}
with $T_{\mu \nu}$ being the energy-momentum tensor for the Maxwell field, defined as
\begin{equation} T_\mu{}^\nu:={ \textsl g}_{\rho
\sigma}{\textsl F}^{^{\nu \rho}}{{{\textsl F}}_\mu}^{\sigma}-\displaystyle{1 \over 4} {\delta_\mu}^{\nu} {\textsl g}^{\lambda \rho} {\textsl g}^{\beta \sigma} {\textsl F}_{\lambda \beta}
{\textsl F}_{\rho \sigma}.
\end{equation}
The trace of  the field equations (\ref{fec}) gives Eq. (\ref{tras1}).

\subsection{Charged solution for Taub-NUT spacetime}
By combining Eq. (\ref{fec}) with the spacetime (\ref{m22}), we get
\begin{eqnarray} \label{fe4}
& &S_t{}^t=\frac{\aleph(k^2N_1s'^2
-kss'[2k'N_1+kN'_1]-2k^2sN_1s'' -8s^3\chi^2+8k^2s^2\Lambda)+8sN_1k^2q'^2+16s^2\chi^2 q^2+8s\csc\theta h_\theta[\csc\theta h_\theta-2q\chi]}{4k^2s^2}=0,\nonumber\\
& &S_r{}^r=\frac{1}{4s^2k^2}\Big(\aleph(k^2N_1s'^2-2k^2sN_1s''-sk^2s'N'_1+2s^2N_1k'^2-4ks^2N_1k''-2ks^2k'N'_1+8k^2s^2\Lambda)+4sN_1k^2q'^2+16s^2\chi^2 q^2\nonumber\\
& &+4s\csc\theta h_\theta[\csc\theta h_\theta-4q\chi]\Big)
=0,\nonumber\\
 && S_\theta{}^\theta=S_\phi{}^\phi= \frac{\aleph(2skN_1k''+kk'(sN_1)'-8s^2\chi^2-4ks[1+2k\Lambda])+8sN_1k^2q'^2+16s\chi^2 q^2+4s\csc\theta h_\theta[\csc\theta h_\theta-4q\chi]}{4sk^2}=0,\nonumber\\
& &S_\phi{}^t= \frac{\chi\cos\theta}{2k^2s^2}\Big[\aleph\{(2ks^2N_1k''-2k^2sN_1s''+k^2N_1s'^2-skk'(s'N_1-sN'_1)-k^2ss'N'_1-4s^2(2\chi^2s+k))\}+4N_1k^2q'^2+16s^2\chi^2 q^2\Big]\nonumber\\
& &+4s^2\csc\theta h_\theta[\csc\theta h_\theta-4q\chi]=0,
 \end{eqnarray}
where Eq. (\ref{tras}) is used and we assume  the vector potential has the form
\begin{equation} \label{pot1}  A=q(r)dt^2+[2\chi q(r)\cos(\theta)+h(\theta)] d\phi.
\end{equation}
Equation (\ref{pot1}) shows that the vector potential consists of the electric field as well as the magnetic field.
The solution of Eq. (\ref{fe4}) takes the form
\begin{eqnarray} \label{s44}
& &k(r)=\varrho^2,\qquad \qquad s(r)=s_1(r)N_1(r), \qquad \qquad \textrm where \qquad \qquad s_1(r)=\frac{r^2}{c_3\varrho^2-\chi^2},\nonumber\\
 & &N_1(r)=\frac{c_3\varrho^2-\chi^2}{3c_3{}^3r^2\varrho^2\aleph}\Big(3c_3{}^4[c_{4}{}^2+4c_{5}{}^2\chi^2]+3c_3{}^3[4c_{6}c_{5}\chi-c_{7}\aleph\sqrt{c_3{}^2
\varrho-\chi^2}]+c_3{}^2\Big[2\varrho^2\Lambda (r^2-24\beta+\chi^2)-32\beta \varrho^4\Lambda^2+3(c_{6}{}^2+\varrho^2)\Big]\nonumber\\
 & &+2c_3\chi^2\aleph[4\Lambda \varrho^2-3]-16\chi^4\Lambda \aleph\Big), \qquad  h(\theta)=c_{6}\cos\theta, \qquad   q(r)=\frac{1}{c_3\varrho^2}[c_{5}c_3(c_3\varrho^2 -2\chi^2)+c_3c_{4}\sqrt{c_3\varrho^2 -\chi^2}-c_{6}\chi].\nonumber\\
 & & \end{eqnarray}
  The horizons of solution (\ref{s44}) are plotted in figure \ref{Fig:1}\subref{fig:1b} which shows two horizons, one for the event and the second is the cosmological horizons.
  Using  Eq. (\ref{m22})  then metric spacetime of solution (\ref{s44}) takes the form
\begin{eqnarray} \label{metr44}
ds^2&=&-\Big\{\frac{1}{3c_3{}^3\varrho^2\aleph}\Big(3c_3{}^4[c_{4}{}^2+4c_{5}{}^2\chi^2]+3c_3{}^3[4c_{6}c_{5}\chi+c_{7}\aleph\sqrt{c_3{}^2
\varrho-\chi^2}]+c_3{}^2\Big[2\varrho^2\Lambda (r^2-24\beta+\chi^2)-32\beta \varrho^4\Lambda^2\nonumber\\
 & &+3(c_{6}{}^2+r^2+\chi^2)\Big]+2c_3\chi^2\aleph[4\Lambda \varrho^2-3]-16\chi^4\Lambda \aleph\Big)\Big\}dt^2+\varrho^2d\theta^2+\Bigg\{\frac{c_3\varrho^2-\chi^2}{3c_3{}^3r^2\varrho^2\aleph}\Big(3c_3{}^4[c_{4}{}^2+4c_{5}{}^2\chi^2]
 +3c_3{}^3[4c_{6}c_{5}\chi
\nonumber\\
& &-c_{7}\aleph\sqrt{c_3{}^2
\varrho-\chi^2}]+c_3{}^2\Big[2\varrho^2\Lambda (r^2-24\beta+\chi^2)-32\beta \varrho^4\Lambda^2+3(c_{6}{}^2+r^2+\chi^2)\Big]+2c_3\chi^2\aleph[4\Lambda \varrho^2-3]-16\chi^4\Lambda \aleph\Big)\Bigg\}^{-1}dr^2\nonumber\\
& &-\frac{1}{3c_3{}^3\varrho^2\aleph}\Big[3c_3{}^3\aleph \varrho^4 \sin^2\theta-12c_{7}c_3{}^3 \aleph \chi^2 \cos^2\theta\sqrt{c_3\varrho^2-\chi^2}-4\chi^2\cos^2\theta(3c_3{}^4[4\chi^2c_{5}^2+c_{4}^2]+12c_3{}^3c_{5}c_{6}\chi\nonumber\\
& &+c_3{}^2
[2\aleph\chi^4\Lambda+\chi^2\aleph (3+4r^2\Lambda)-32\beta r^4\Lambda^2+2r^2\Lambda (r^2-24\beta)+3(r^2+c_{6})]+2c_3\aleph \chi^2[4\Lambda\varrho^2-3]
-16\chi^4\aleph \Lambda)\Big]d\phi^2\nonumber\\
& &+\frac{4\chi\cos\theta}{3c_3{}^3\varrho^2\aleph}\Big[3c_{7}c_3{}^3 \aleph\sqrt{c_3\varrho^2-\chi^2}+3c_3{}^4[4\chi^2c_{5}^2+c_{4}^2]+12c_3{}^3c_{5}c_{6}\chi+c_3{}^2
[2\aleph\chi^4\Lambda+\chi^2\aleph (3+4r^2\Lambda)-32\beta r^4\Lambda^2\nonumber\\
& &+2r^2\Lambda (r^2-24\beta)+3(r^2+c_{6})]+2c_3\aleph \chi^2[4\Lambda\varrho^2-3]
-16\chi^4\aleph \Lambda\Big]d\phi dt\;.\nonumber\\
& &\end{eqnarray}
From Eq. (\ref{metr44}), we can see that solution (\ref{s44}) behaves asymptotically as AdS/dS and  informs us that it is a new solution that depends on the dimension parameter $\beta$ that must satisfies \[\aleph\neq 0\Rightarrow \beta\neq \frac{1}{16 \Lambda}.\]
Equation (\ref{metr44}) is singular at $\varrho=0$ and  $r=0$. Moreover, the metric (\ref{metr44}) has another singularity at \begin{eqnarray} \label{scal444}
&&3c_3{}^4[c_{4}{}^2+4c_{5}{}^2\chi^2]
 +3c_3{}^3[4c_{6}c_{5}\chi-c_{7}\aleph\sqrt{c_3{}^2
\varrho-\chi^2}]+c_3{}^2\Big[2\varrho^2\Lambda (r^2-24\beta+\chi^2)-32\beta \varrho^4\Lambda^2+3(c_{6}{}^2+r^2+\chi^2)\Big]\nonumber\\
&&+2c_3\chi^2\aleph[4\Lambda \varrho^2-3]-16\chi^4\Lambda \aleph=0.\end{eqnarray}
Equation (\ref{scal444}) is a fourth order algebraic equation that has at least two real positive roots.

The  Kretschmann invariant, squared Ricci and Maxwell field  of solution (\ref{s44}) take the form
 \begin{eqnarray} \label{scal44} && R^{\mu \nu \lambda \rho}R_{\mu \nu \lambda \rho}=\frac{\mathbb{F}_2(r)}{3\aleph^2\varrho^{12}}, \qquad \qquad R^{\mu \nu}R_{\mu \nu}=\frac{\mathbb{F}_3(r)}{\aleph^2\varrho^8}, \qquad \qquad F^{\mu \nu}F_{\mu \nu}=\frac{\mathbb{F}_4(r)}{\varrho^8},
\end{eqnarray}
 where $\mathbb{F}_2(r)$, $\mathbb{F}_3(r)$ and $\mathbb{F}_4(r)$ are  polynomial functions and we have put $c_3=1$.
 It is understood from Eq. (\ref{scal44}) that a singularity exists at $\varrho=0 \Rightarrow r=0$ when $\chi=0$.  It is important to mention that the singularity that arise from Eq. (\ref{scal444}) and makes the metric (\ref{metr44}) singular does not make the  Kretschmann invariant and squared Ricci of Eq. (\ref{scal44}) divergent.
\section{Physical properties of the  black hole solutions}\label{S4}
We investigate the Taub--NUT non-charged case to the metric (\ref{metr3}) which can be rewritten as\footnote{The family of the Taub--NUT spacetimes is given by
$$ds^2=-\Delta(r)dt^2+\varrho^2d\theta^2
+\frac{dr^2}{\Delta(r)}+\left\{\varrho^2\sin^2\theta-4\chi^2\, \Delta(r)\,[\cos \theta+C]^2\right\}d\phi^2-4\chi\,\Delta(r)\, [\cos \theta+C] drdt,$$ where C is an additional parameter related to the  large coordinate transformation $t\rightarrow t+C\phi$. Note that C should be considered as physical rather than pure gauge parameter, since it changes the asymptotic behavior of the metric. Its introduction was often used to modify the position of the Misner string: for $C=-1$ it lies at the southern hemisphere, for $C=1$ at the northern and for $C=0$ at both of them. Equation (\ref{metrp3}) belong to $C=0$ Taub-NUT family and in that case Minser string lies on both of the northern and southern hemispheres \cite{Clement:2015cxa}.}
\begin{eqnarray} \label{metrp3}
&&ds^2=-\Delta_1(r)dt^2+\varrho^2d\theta^2
+\frac{dr^2}{\Delta_1(r)}+\left\{\varrho^2\sin^2\theta-4\chi^2\, \Delta_1(r)\,\cos^2 \theta\right\}d\phi^2+4\chi\,\Delta_1(r)\, \cos\theta\, d\phi\, dt\;,\end{eqnarray}
 where $c_2=-2m$ and $$\Delta_1(r)=1+\frac{2\Lambda\varrho^2}{3}+\frac{2\chi^2[4\Lambda\varrho^2-3]}{3\varrho^2}-\frac{16\chi^4\Lambda}{3\varrho^2}-
 \frac{2mr}{\varrho^2}= 0.$$  In the limiting case when $\chi\rightarrow 0$  Eq. (\ref{metrp3}) reduces to Schwarzschild AdS/dS spacetime \cite{Nashed:2013bfa}.
The number of the horizons of Eq. (\ref{metrp3}) are the  positive real roots of
$\Delta_1(r)=0$ which has four roots, two of them are real that represent the inner and outer horizons. The metric (\ref{metrp3}) is singular at $\varrho=0 \rightarrow$ $r=0$ and $\chi=0$ in addition to $\Delta_1(r)=0 .$  If we want to overcome the problem  encountered in the coordinate $(t,r,\theta,\phi)$
system, singularity at $\Delta_1(r)=0$, an obvious place to start is the time-coordinate $t$. We will replace $t$ by a
coordinate $(v,r,\theta,\phi)$, where $v=t+r_\ast$ and $r_\ast$ is defined by
\begin{eqnarray} \label{ref1}
dr_\ast=\Delta_1{}^{-1}(r) dr\,.
\end{eqnarray}
In terms of the new coordinates the line-element (\ref{metrp3}) becomes
\begin{eqnarray} \label{metrp33}
&&ds^2=-\Delta_1(r)dv^2+2dvdr+\varrho^2d\theta^2
+\left\{\varrho^2\sin^2\theta-4\chi^2\, \Delta_1(r)\,\cos^2 \theta\right\}d\phi^2+4\chi\,\cos\theta[\Delta_1(r)dv-dr] d\phi\;.\end{eqnarray}
Equation (\ref{metrp33}) shows that there are no more factors of $\Delta_1(r)$ in the denominator, and the metric is regular at the inner and outer horizons. The only remaining singularity is the curvature singularity at $\varrho=0$.

For the charged Taub-NUT  the metric (\ref{metr44}) can be rewritten as
\begin{eqnarray} \label{metrp4}
&&ds^2=-\Delta_2(r)dt^2+\varrho^2d\theta^2
+\frac{dr^2}{\Delta_2(r)}+\left\{\varrho^2\sin^2\theta-4\chi^2\, \Delta_2(r)\,\cos^2 \theta \right\}d\phi^2+4\chi\,\Delta_2(r)\, \cos\theta drdt\;,\end{eqnarray}
 where we have used $c_3=1$, $c_{4}=q/2$, $c_{5}=q_1$, $c_{6}=q/2$ and $c_{7}=-2m$  to ensure a flat spacetime when $\chi\rightarrow0$, $\beta\rightarrow0$ and $\Lambda\rightarrow0$. Here $\Delta_2$ is defined as
    $$\Delta_2(r)=\frac{1}{3\varrho^2\aleph}\Big(3[q^2+4q_{1}{}^2\chi^2]+3[2q_1q\chi-2mr \aleph]+2\varrho^2\Lambda (\varrho^2-24\beta)-32\beta \varrho^4\Lambda^2+3\varrho^2-16\chi^4\Lambda \aleph+2\chi^2\aleph[4\Lambda \varrho^2-3]\Big).$$

  In the limiting case $\chi\rightarrow 0$ and $\beta\rightarrow0$, Eq. (\ref{metrp4}) reduces to Reissner-Nordstr\"om AdS/dS spacetime \cite{Nashed:2013bfa}. The most interesting thing is the fact that the charge $q_1$ is accompanied with the dimension parameter $\chi$. So, if $\chi=0$ then the charge $q_1$ will disappear however, the inverse is not correct, i.e., when $q_1\rightarrow 0$ then $\chi$ will not disappear.   As usual the number of horizons of the spacetime  (\ref{metrp4}) are the positive roots of $$\Delta_2(r)=3[q^2+4q_{1}{}^2\chi^2]+3[2q_1q\chi-2mr \aleph]+2\varrho^2\Lambda (\varrho^2-24\beta)-32\beta \varrho^4\Lambda^2+3\varrho^2-16\chi^4\Lambda \aleph+2\chi^2\aleph[4\Lambda \varrho^2-3]=0,$$  which has two real roots. Using the procedure applied in the neutral spacetime we can show that the line-element (\ref{metrp4}) can has the form
 \begin{eqnarray} \label{metrp44}
&&ds^2=-\Delta_2(r)dv^2+2dvdr+\varrho^2d\theta^2
+\left\{\varrho^2\sin^2\theta-4\chi^2\, \Delta_2(r)\,\cos^2 \theta\right\}d\phi^2+4\chi\,\cos\theta[\Delta_2(r)dv-dr] d\phi\;,\end{eqnarray} where $v$ in this case has the same form given in the neutral case and $r_\ast$ is defined by
\begin{eqnarray} \label{ref1}
dr_\ast=\Delta_2{}^{-1}(r) dr\,.\end{eqnarray}

Now we consider the energy conditions that have the following constrains   \cite{Nashed:2018piz}:
 \begin{eqnarray} \label{ec}
\diamond  & &SEC : \rho+p_r\geq0, \qquad \qquad \rho+p_t\geq0, \qquad \qquad \rho+p_r+2p_t\geq0, \nonumber\\
 \diamond  &&WEC : \rho\geq0, \qquad \qquad \; \; \;\;\rho+p_r\geq0,\qquad \qquad\rho+p_t\geq0, \nonumber\\
 \diamond  &&NEC : \rho+p_r\geq0,\qquad \;\;\; \rho+p_t\geq0, \nonumber\\
 \diamond  && DEC :  \rho>\geq0,\qquad \qquad \rho\pm p_r\geq0, \qquad \qquad \rho\pm p_t\geq0,
\end{eqnarray}
where $T_0{}^0=\rho$ is the density, $T_1{}^1=p_r$ is the radial pressure, and $T_2{}^2=T_3{}^3=p_t$ is the tangential pressure.
Straightforward calculations of charged Taub-NUT black hole solution (\ref{s44}) gives
\begin{eqnarray} \label{cal}
\diamondsuit & &\mathrm{Strong~ Energy ~Condition} : \rho+p_r= \frac{2(2qq_1\chi-2q_1{}^2\chi^2-q^2)}{\varrho^4}> 0, \quad \rho+p_r+2p_t=0, \nonumber\\
 \diamondsuit &&\mathrm{Weak~ Energy ~Condition} : \rho=\frac{2qq_1\chi-2q_1{}^2\chi^2-q^2}{\varrho^4}> 0, \quad \rho+p_r=\frac{2(2qq_1\chi-2q_1{}^2\chi^2-q^2)}{\varrho^4}> 0,\quad \rho+p_t= 0, \nonumber\\
 \diamondsuit &&\mathrm{Null~ Energy ~Condition} : \rho+p_r=\frac{2(2qq_1\chi-2q_1{}^2\chi^2-q^2)}{\varrho^4}>  0, \quad \rho+p_t=0, \nonumber\\[10pt]
\diamondsuit &&\mathrm{Dominant~ Energy~ Condition} : \rho=\frac{2qq_1\chi-2q_1{}^2\chi^2-q^2}{\varrho^4}>0, \quad \rho- p_r=0, \quad \; \; \; \rho+ p_t=0.
\end{eqnarray}
 Hence, all of the energy conditions are satisfied provided that $qq_1>\frac{2q_1{}^2\chi^2+q^2}{2\chi}$. We have shown that the charged Taub-NUT black hole configuration  may fulfil the energy conditions.

\section{Thermodynamics of the derived black holes}\label{S5}
\begin{figure}
\centering
\subfigure[~The Taub-NUT non-charged case]{\label{fig:1a}\includegraphics[scale=0.35]{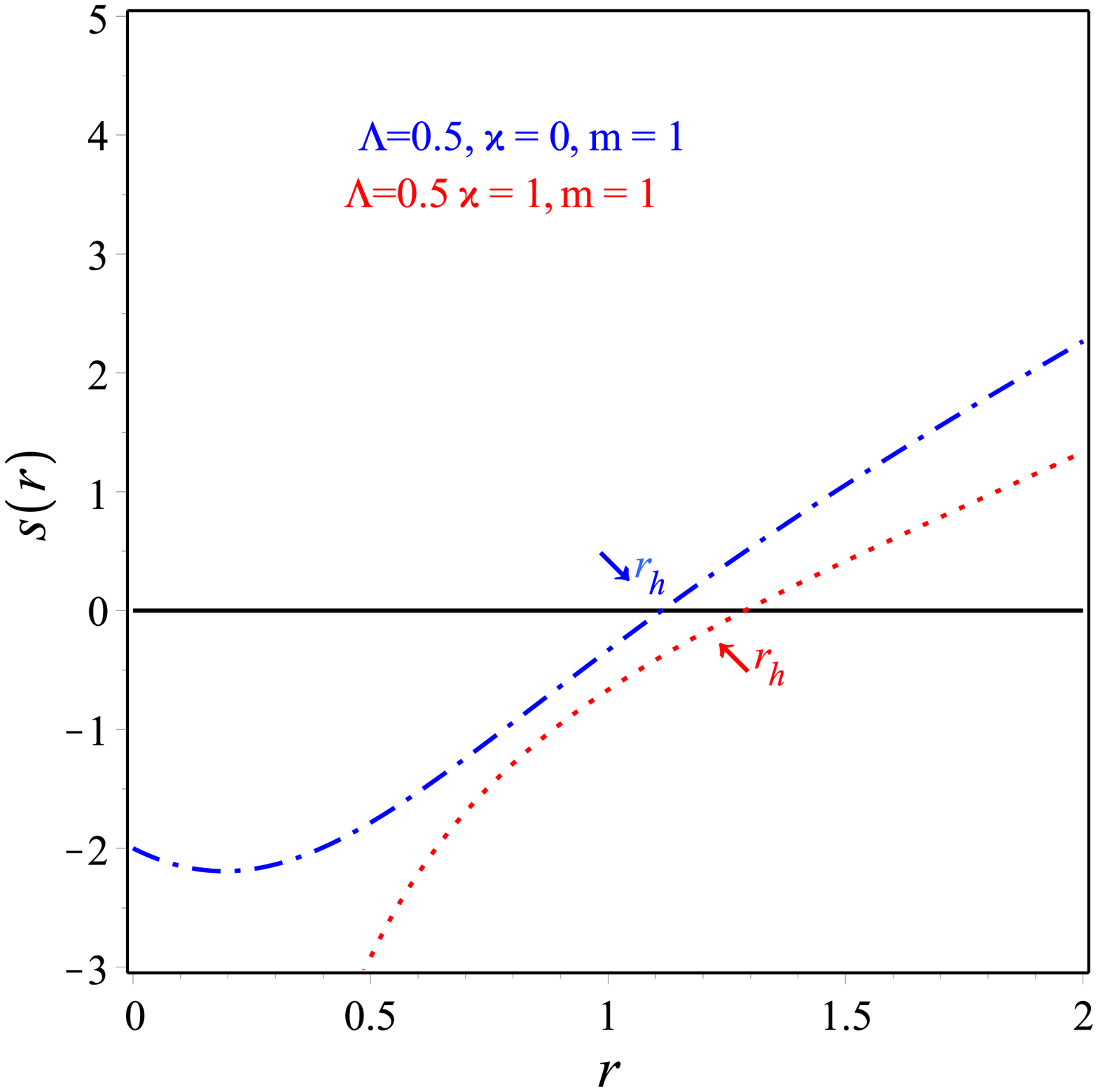}}\hspace{0.2cm}
\subfigure[~The Taub-NUT charged case]{\label{fig:1b}\includegraphics[scale=0.35]{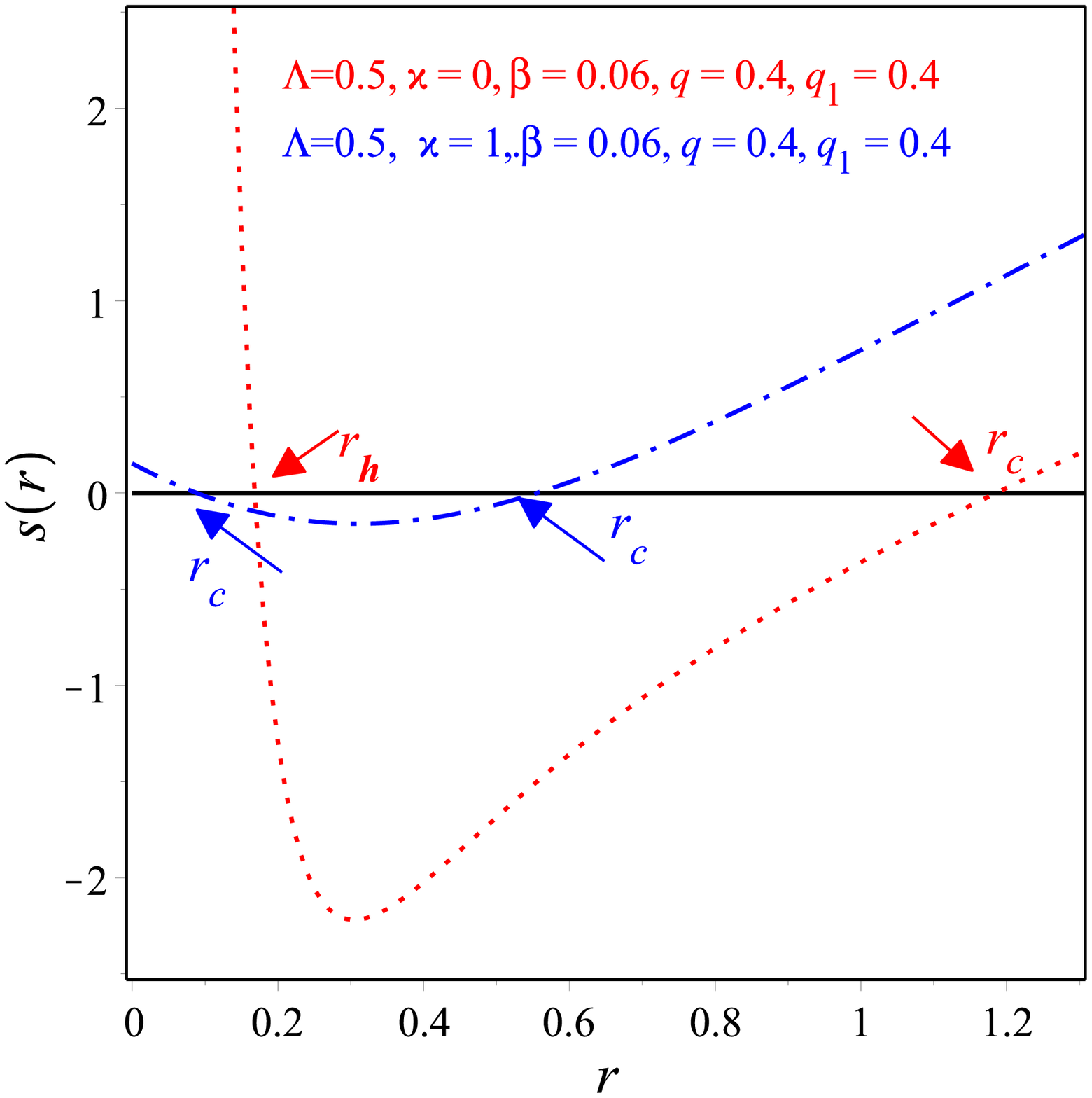}}
\caption{Schematic plots of $N(r)$ and $s(r)$ that characterize the horizons of the black holes by setting $N(r)=0$ and $s(r)=0$: \subref{fig:1a} For the Taub-NUT non-charged case, the function $s(r)$ is given by (\ref{s3}); \subref{fig:1b} For the Taub-NUT charged case, the function $s(r)$ is given by (\ref{s44}).}
\label{Fig:1}
\end{figure}

To study different thermodynamical properties \cite{Hunter:1998qe,Hawking:1998ct,Bekenstein:1972tm,Bekenstein:1973ur,Gibbons:1977mu} of the black hole solutions given by  Eqs.  (\ref{s3}) and (\ref{s44}), we start by obtaining roots for $N(r) = 0$ as well as  $s(r) = 0$. These horizons can be seen in Figs. \ref{Fig:1}\subref{fig:1a} and \ref{Fig:1}\subref{fig:1b}.

The Bekenstein-Hawking entropy is represented as
\begin{equation}\label{ent}
S(r_h)=\frac{1}{4}A=\pi r_h{}^2,
\end{equation}
where $A$ is the area of the  black hole horizon. The stability of the black hole thermodynamics of Eqs.  (\ref{s3}) and (\ref{s44}) can be tested  by exploring the heat capacity, $C_h$, which is defined as \cite{Nouicer:2007pu,e13121967}
\begin{equation}\label{m55}
C_h=\frac{\partial m}{\partial T}\equiv \frac{\partial m}{\partial r_h} \left(\frac{\partial r_h}{\partial T}\right).
\end{equation}
If $C_{h} > 0$, thermodynamics for black holes is stable. On the other hand, for $C_h< 0$, it is unstable. To understand this, we assume that at some point and due to thermal fluctuations, the black hole absorbs more radiation than it emits which makes the heat capacity positive. On the contrary, when the black hole emits more radiation than it absorbs, the heat capacity becomes negative. Thus, the black holes with negative heat capacities are thermally unstable.

\begin{figure}
\centering
\subfigure[~Mass within horizon $r_h$ Taub-NUT non-charged case]{\label{fig:2a}\includegraphics[scale=0.35]{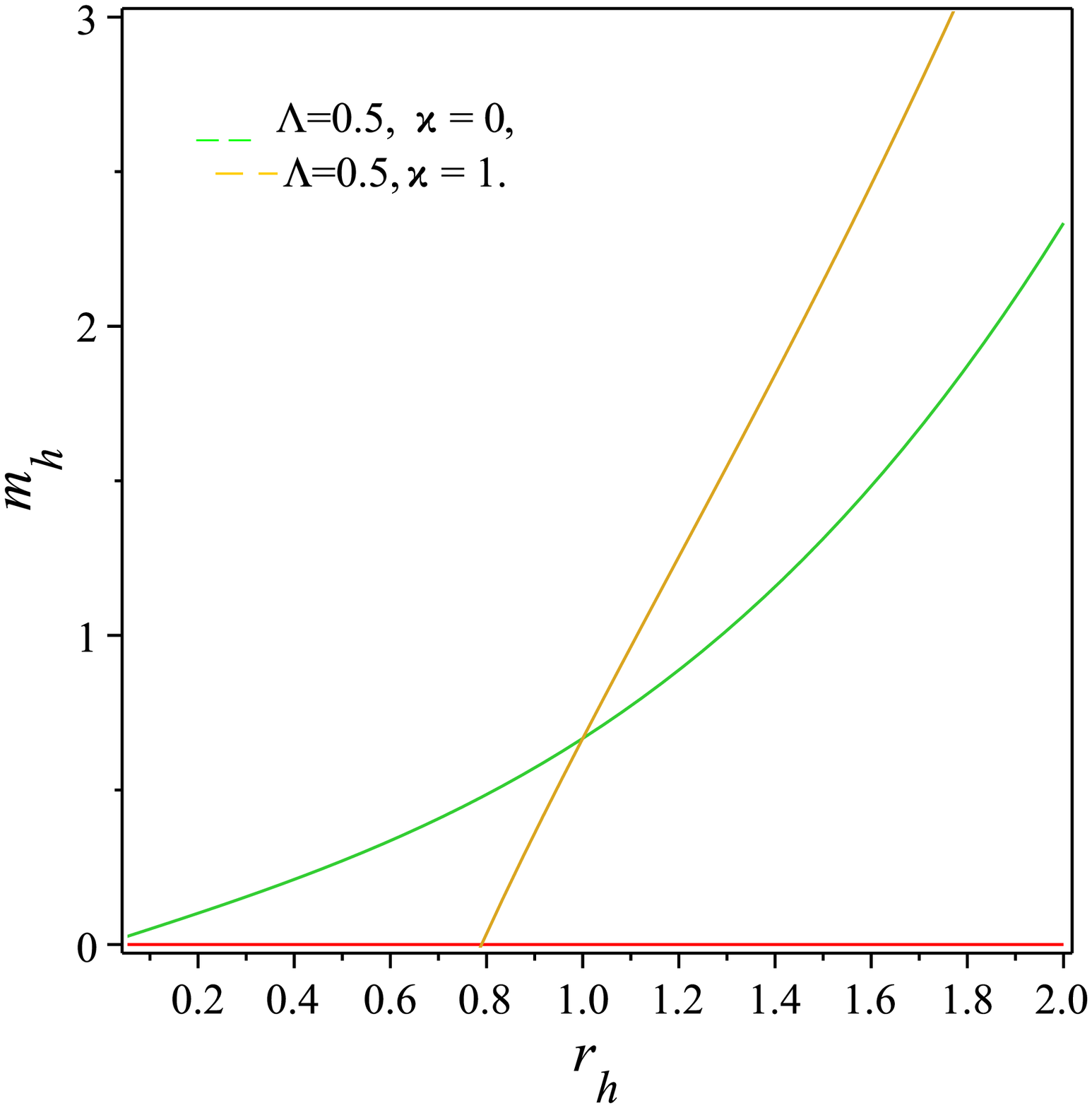}}\hspace{0.2cm}
\subfigure[~Mass within horizon $r_h$ Taub-NUT charged case]{\label{fig:2b}\includegraphics[scale=0.35]{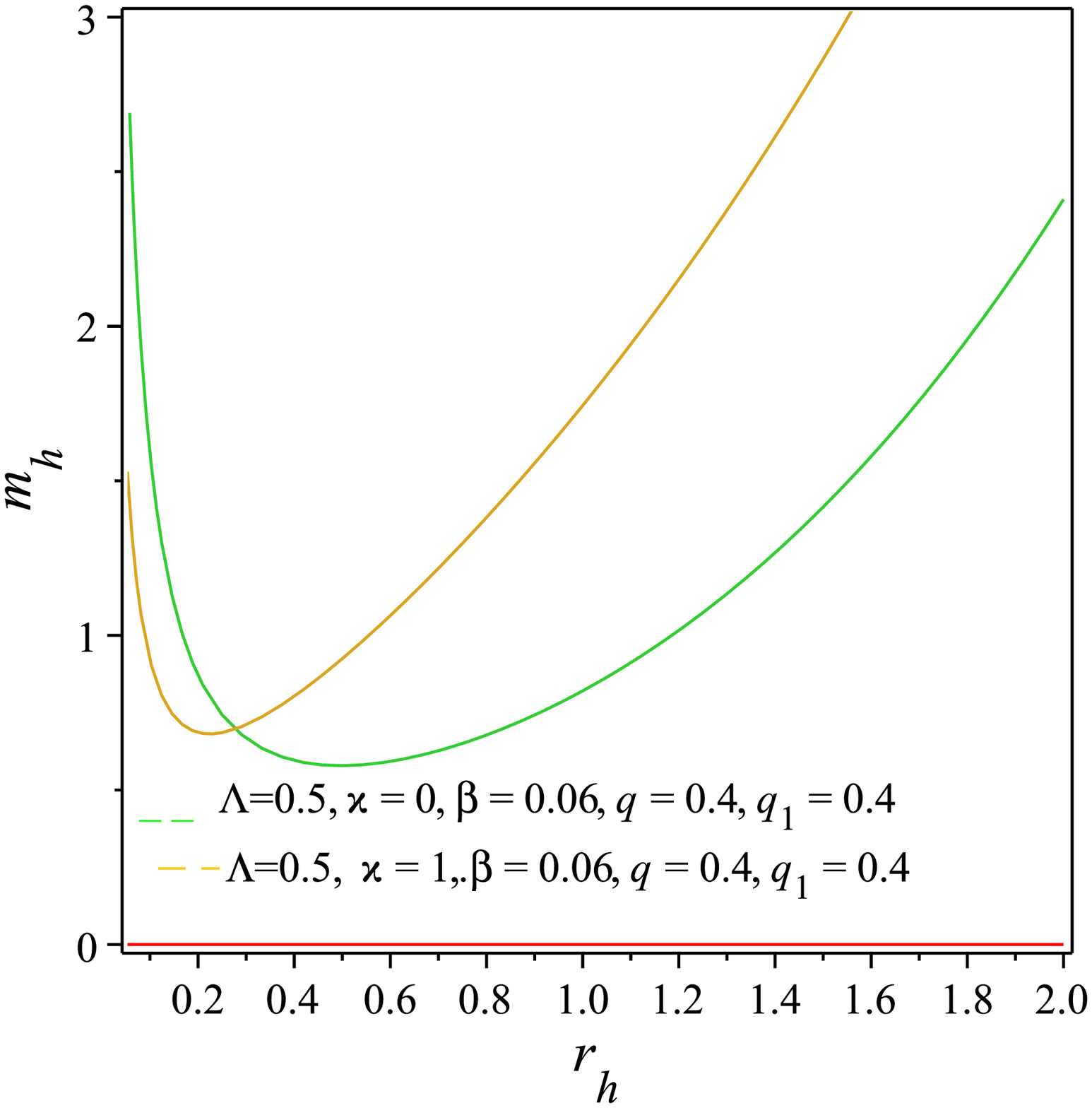}}
\caption{Mass inside the horizon $r_h$ for Eqs.  (\ref{s3}) and (\ref{s44}).}
\label{Fig:2}
\end{figure}

In order to evaluate Eq. (\ref{m55}) we must calculate the black hole mass within the inner horizon $r_h$. To this end we set $N(r_h) = 0$, for Taub-NUT and charged Taub-NUT spacetimes, and $s(r_h)=0$ , for Taub-NUT and charged Taub-NUT spacetimes, then we get
\begin{eqnarray} \label{m33}
&&{m_h}_{{}_{{}_{{}_{{}_{\tiny Eq. (\ref{metrp3})}}}}}=\frac{1}{6r_h}\left[2\Lambda r_h{}^4+3 r_h{}^2(1+4\chi^2\Lambda)-3\chi^2(1+2\chi^2\Lambda) \right], \nonumber\\
&&{m_+}_{{}_{{}_{{}_{{}_{\tiny Eq. (\ref{metrp4})}}}}}\frac{1}{6r_h\aleph}\left[3(\varrho_h{}^2+q^2)-6\chi^2+6q_1(q+2q_1)-2\Lambda(3\chi^4+24\beta r_h{}^2-6\chi^2[r_h{}^2+4\beta]-r_h{}^4)-32\beta \Lambda \{r_h{}^4+6\chi^2 r_h{}^2-3\chi^4\}\right],\nonumber\\
&&
\end{eqnarray}
where $\varrho_h=\sqrt{r_h{}^2+\chi^2}$.
We plot the black hole mass within the radius of the horizon $r_h$ in Fig. \ref{Fig:2}. The black hole size varies between the inner $r_{h}$ and cosmological $r_c$ horizons \cite{Fernando:2016ksb,Katsuragawa:2014hda}.

\begin{figure}
\centering
\subfigure[~Temperature at cosmological horizon Taub-NUT non-charged case]{\label{fig:3a}\includegraphics[scale=0.35]{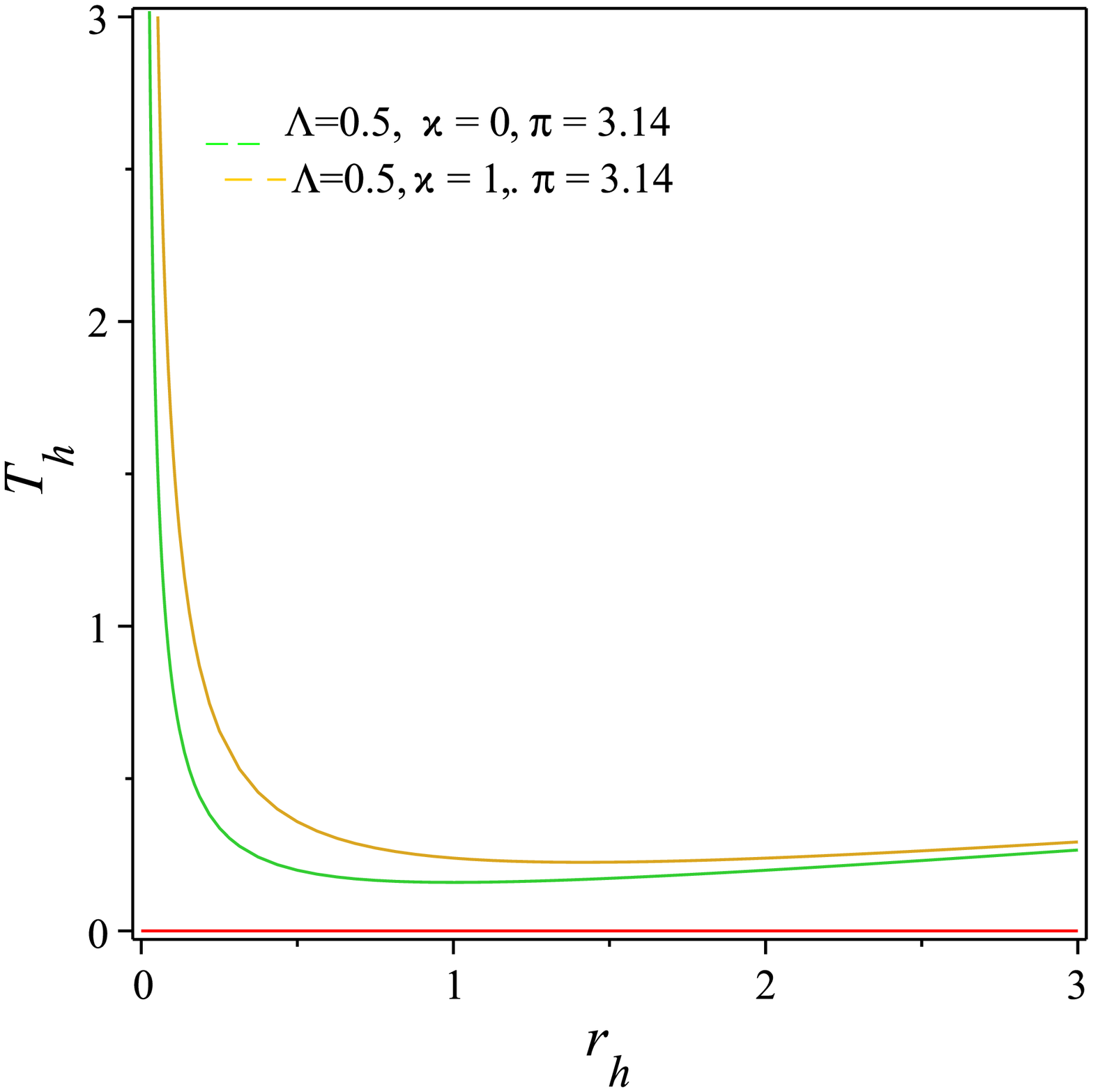}}\hspace{0.2cm}
\subfigure[~Temperature at cosmological horizon Taub-NUT charged case]{\label{fig:3b}\includegraphics[scale=0.35]{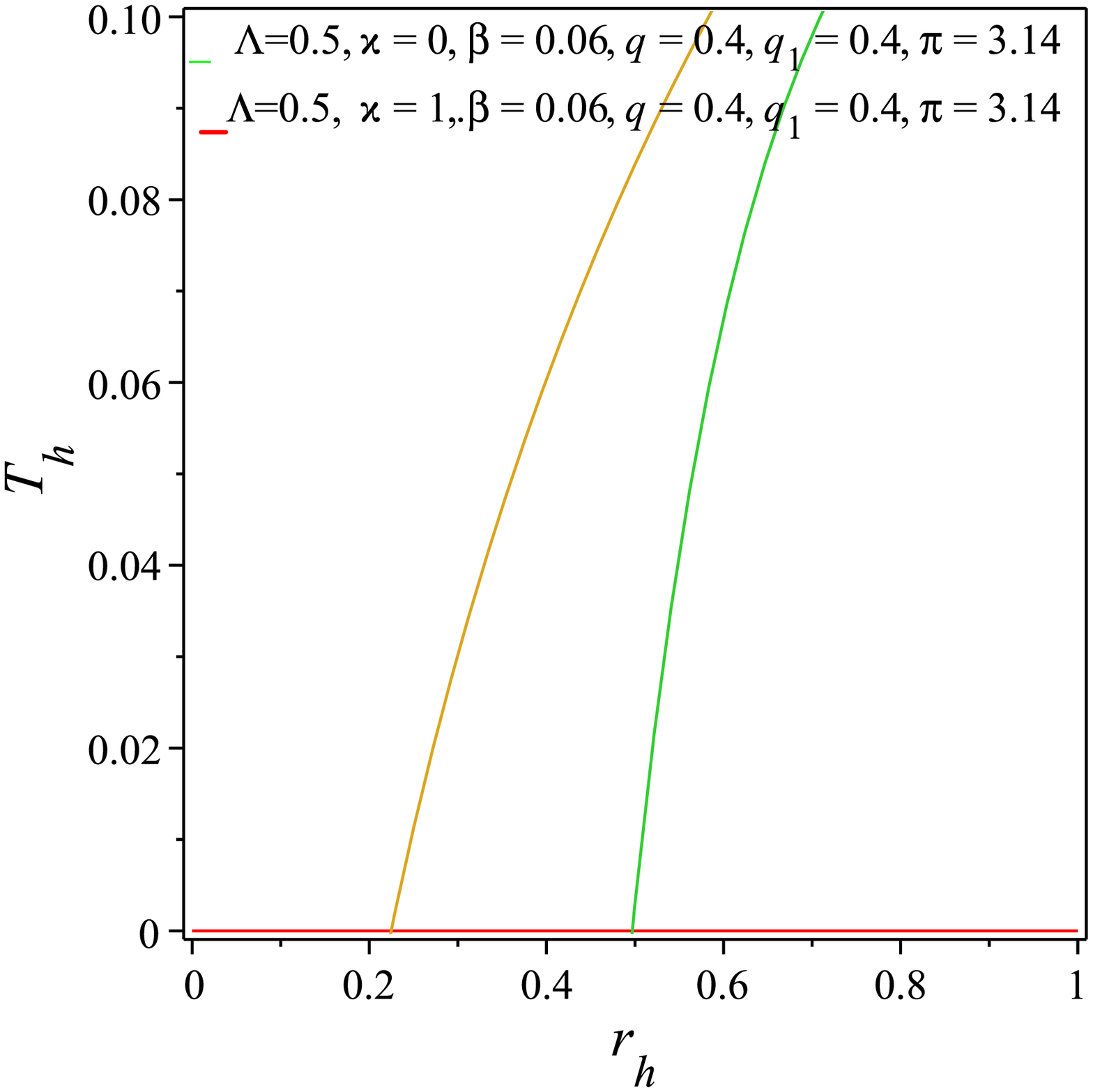}}
\caption{Temperature at the cosmological horizon for Eqs. (\ref{s3}) and (\ref{s44}).}
\label{Fig:3}
\end{figure}

The black hole Hawking temperature is acquired by the request that there is no singularity at the Euclidean horizon. While, the temperature at the a black hole horizon $r = r_h$ is given by \cite{hawking1975}
\begin{equation}
T = \frac{\kappa}{2\pi}, \qquad \textmd{where} \;  \kappa \; \textmd{ is the surface gravity, represented as } \qquad \kappa= \frac{N'(r_h)}{2}= \frac{s'(r_h)}{2}.
\end{equation}
The Hawking temperatures for (\ref{s3}) and (\ref{s44}) are expressed as
\begin{eqnarray} \label{m44}
{T_h}_{{}_{{}_{{}_{{}_{\tiny Eq. (\ref{metrp3})}}}}}&=&\frac{1}{4\pi r_h}\left[2\Lambda \varrho_h{}^2+1\right], \nonumber\\
{T_h}_{{}_{{}_{{}_{{}_{\tiny Eq. (\ref{metrp4})}}}}}&=&\frac{1}{6r_h\aleph}\left[3[\varrho_h{}^2+q^2]-6\chi[\chi-2\chi q_1{}^2-qq_1]-32\beta\Lambda^2(r_h{}^4+6r_h{}^2\chi^2-3\chi^4)-\Lambda(6\chi^4-12\chi^2[r_h{}^2-4\beta]+2r_h{}^2[24\beta-r_h{}^2])\right]\;,\nonumber\\
& &
\end{eqnarray}
where ${T_h}$ is the Hawking temperature for the cosmological horizon.
We plot the temperatures $T_h$ in Fig. \ref{Fig:3}. We show that $r_{min}$ at which $T_h$ vanishes for the NUT case, but that the ultra-cold black holes are considered for $r_h< r_{min}$.
With the effect of gravity for thermal radiation, in a very-high temperature $T_{max}$, thermal radiation becomes unstable, so that it could collapse to black holes \cite{hawking1983}. As a result, only for $T < T_{max}$, the solution for the pure AdS is stable. Above $T_{max}$, black holes with those very heavy masses could be stable \cite{hawking1983}.

\begin{figure}
\centering
\subfigure[~Heat capacity of Taub-NUT non-charged case]{\label{fig:4a}\includegraphics[scale=0.35]{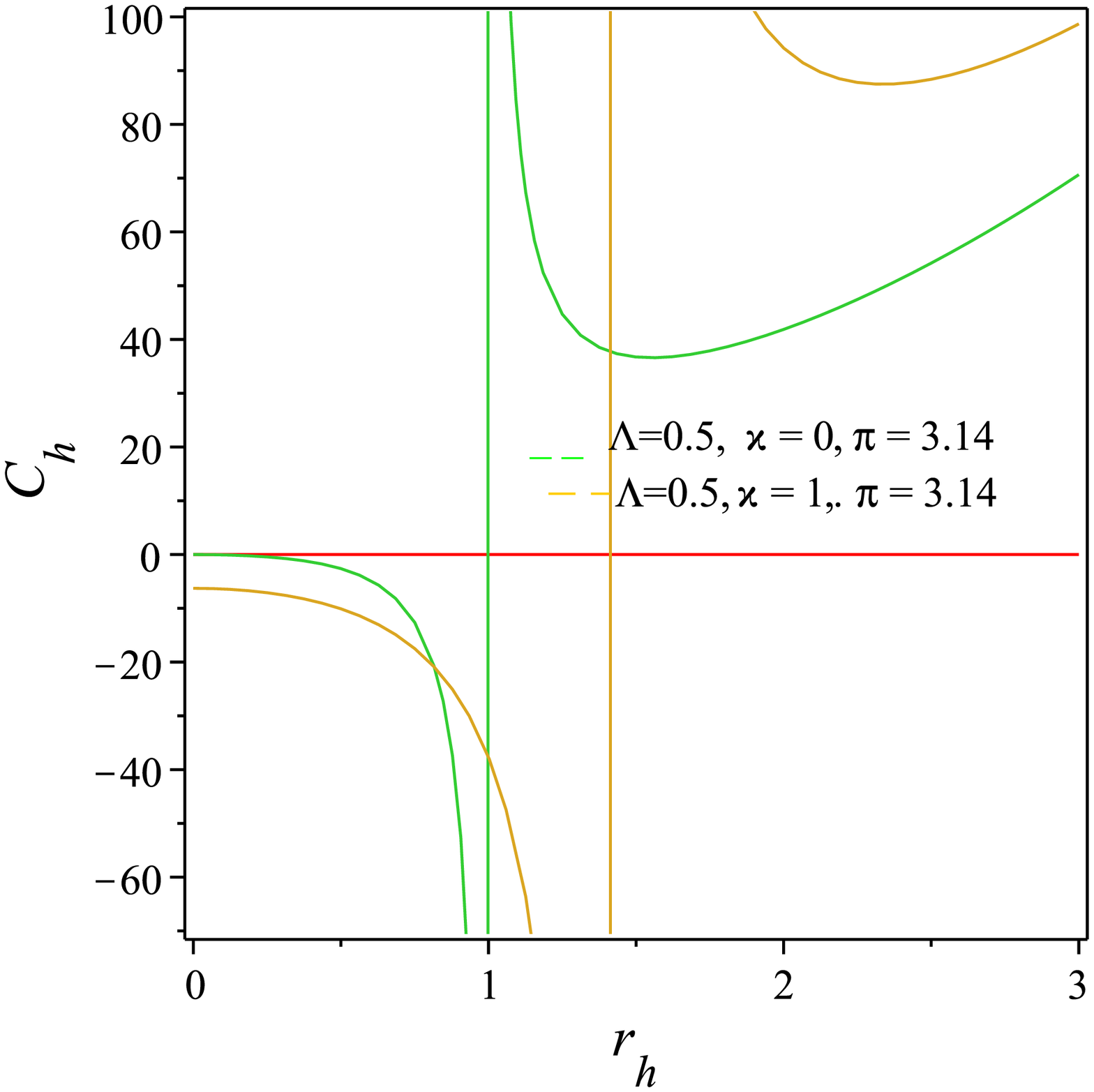}}\hspace{0.2cm}
\subfigure[~Heat capacity of Taub-NUT charged case]{\label{fig:4b}\includegraphics[scale=0.35]{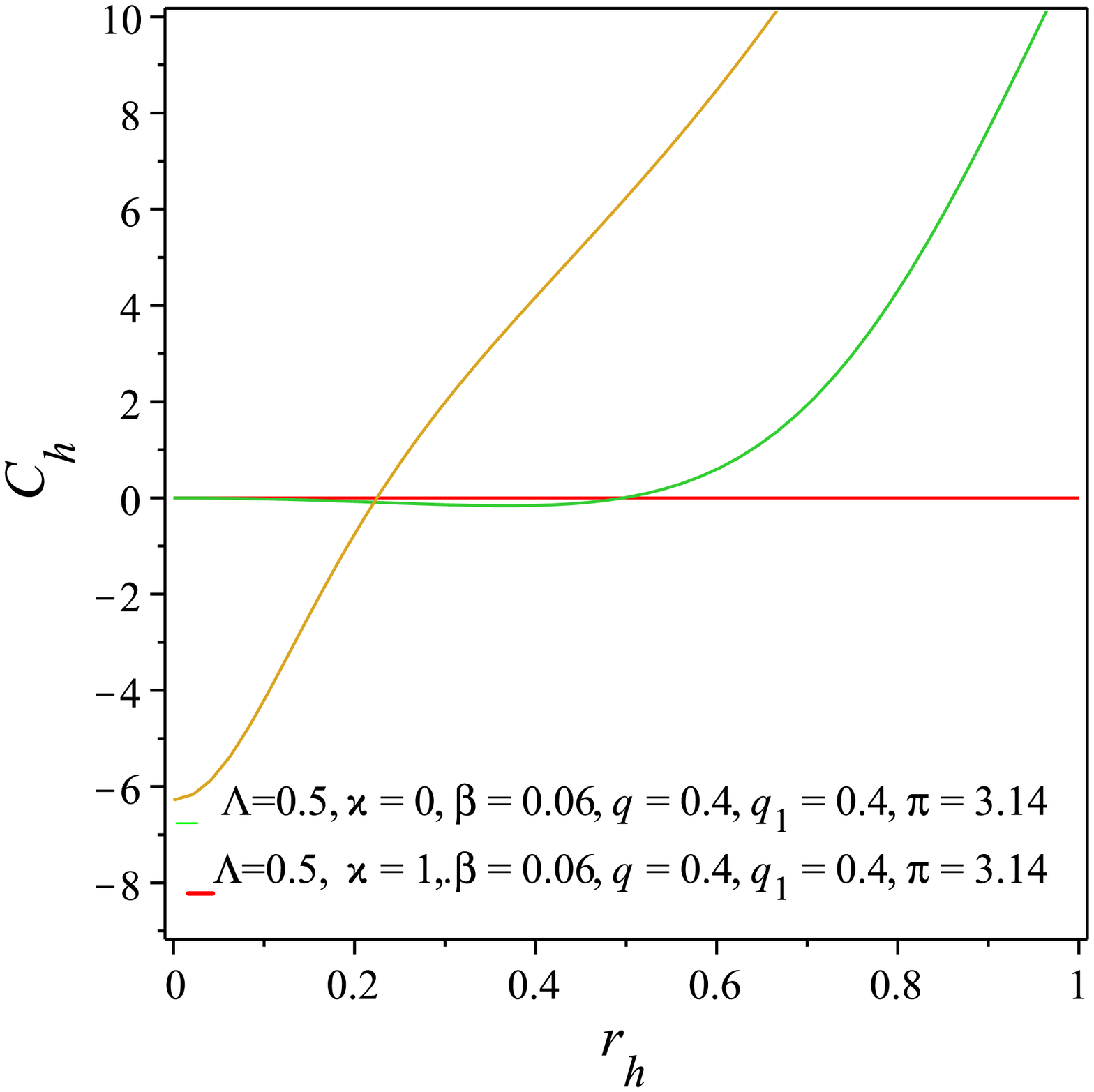}}
\caption{Schematic plots of heat capacity of Eqs. (\ref{s3}) and (\ref{s44}).}
\label{Fig:4}
\end{figure}

We calculate the heat capacity, after substituting Eqs. (\ref{m33}) and (\ref{m44}) into Eq. (\ref{m55}) and get
\begin{eqnarray} \label{m66}
 &&{C_h}_{{}_{{}_{{}_{{}_{\tiny Eq. (\ref{metrp3})}}}}}=\frac{2\pi \varrho_h{}^2(2\Lambda \varrho_h{}^2+1)}{2\Lambda(r_h{}^2-\chi^2)-1},\nonumber\\
\nonumber\\
 &&{C_h}_{{}_{{}_{{}_{{}_{\tiny Eq. (\ref{metrp4})}}}}}=\frac{2\pi [2\chi qq_1-2\Lambda \chi^4\aleph-\chi^2(\aleph[1+4r_h{}^2\Lambda]+4q_1{}^2)-2 \aleph\Lambda r_h{}^4-\varrho_+{}^4(\aleph r_h{}^2-q^2)]}{2\Lambda \chi^6 \aleph+\chi^4[\aleph(2r_h{}^2\Lambda +1)-4q^2]-2qq_1\chi^3-\chi^2(2\Lambda r_h{}^4\aleph-2r_h{}^2[\aleph-6q_1{}^2]+q^2)-6r_h{}^2qq_1+(1-2\Lambda r_h{}^2) r_h{}^4\aleph-3q^2 r_h{}^2}.\nonumber\\
\end{eqnarray}
To have information directly from Eq. (\ref{m66}) is not easy, so we plot them in Fig. \ref{Fig:4} for particular values of the parameters of black holes. For the non-charged case, the heat capacity is negative when $r_h<r_{min}$ at which the temperature has a vanishing value.  For the case $r_h>r_{min}$, the heat capacity become positive, so that the solution can locally be stable. The same conclusion is valid for the charged  Taub-NUT black holes. Note that all the above heat capacity  are characterized by a second-order phase transition \cite{Di96,Hayward:2005gi} as $C_h$ diverges at some critical value $r_h < r_{min}$. In conclusion, Eq. (\ref{m33}) shows that ${\partial m_h}/{\partial r_h}>0$, while the sign for the heat capacity is equal to that for ${\partial T}/{\partial r_h}$. Consequently, we find that $C_h < 0$ when $r_h < r_{min}$ and $C_h > 0$ when $r_h > r_{min}$. In this sense, there are two possible black hole solutions for a given temperature $T > T(r_{min})$, but only the bigger one is thermally stable.

The free energy in grand canonical ensemble also called Gibbs free energy is defined as
\begin{equation} \label{enr}
G(r_h)=M(r_h)-T(r_h)S(r_h).
\end{equation}
Here, $M(r_h)$ is the black hole mass,
$T(r_h)$ is the temperature for the a black hole horizon,
and $S(r_h)$ is the entropy. Using Eqs. (\ref{ent}), (\ref{m33}) and (\ref{m44}) in (\ref{enr}) we get
\begin{eqnarray} \label{m77}
 &&{G_h}_{{}_{{}_{{}_{{}_{\tiny Eq. (\ref{metrp3})}}}}}=\frac{18\Lambda\chi^2r_h{}^2+3r_h{}^2-2\Lambda r_h{}^4-6\chi^2-12\Lambda \chi^4}{12r_h},\nonumber\\
\nonumber\\
 &&{G_h}_{{}_{{}_{{}_{{}_{\tiny Eq. (\ref{metrp4})}}}}}=\frac{1}{12 \aleph \varrho_h{}^2 r_h{}}\Bigg\{6 qq_1[2\chi^2+3r_h{}^2]-6\chi^4[2\Lambda \chi^2\aleph-(\aleph[1+r_h{}^2\Lambda]+4q_1{}^2)]+\chi^2(16\aleph r_h{}^4+3r_h{}^2[\aleph+12q_1{}^2]+6q^2)\nonumber\\
\nonumber\\
&&-r_h{}^2[2 \aleph\Lambda r_h{}^6+3(\aleph r_h{}^2-3q^2)]\Bigg\}.\nonumber\\
\end{eqnarray}

\begin{figure}
\centering
\subfigure[~Free energy of Taub-NUT non-charged case]{\label{fig:5a}\includegraphics[scale=0.35]{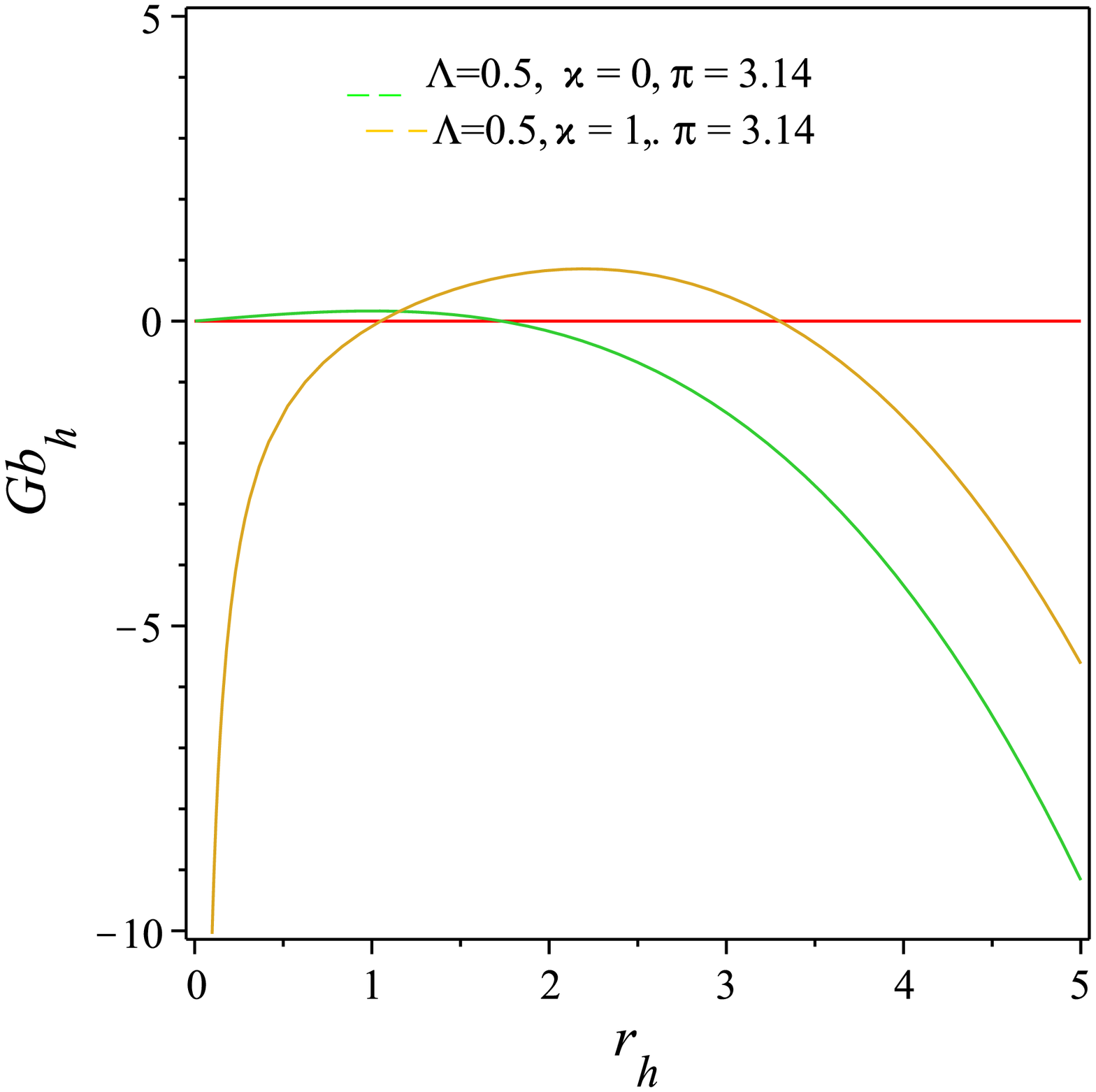}}\hspace{0.2cm}
\subfigure[~Free energy of Taub-NUT charged case]{\label{fig:5b}\includegraphics[scale=0.35]{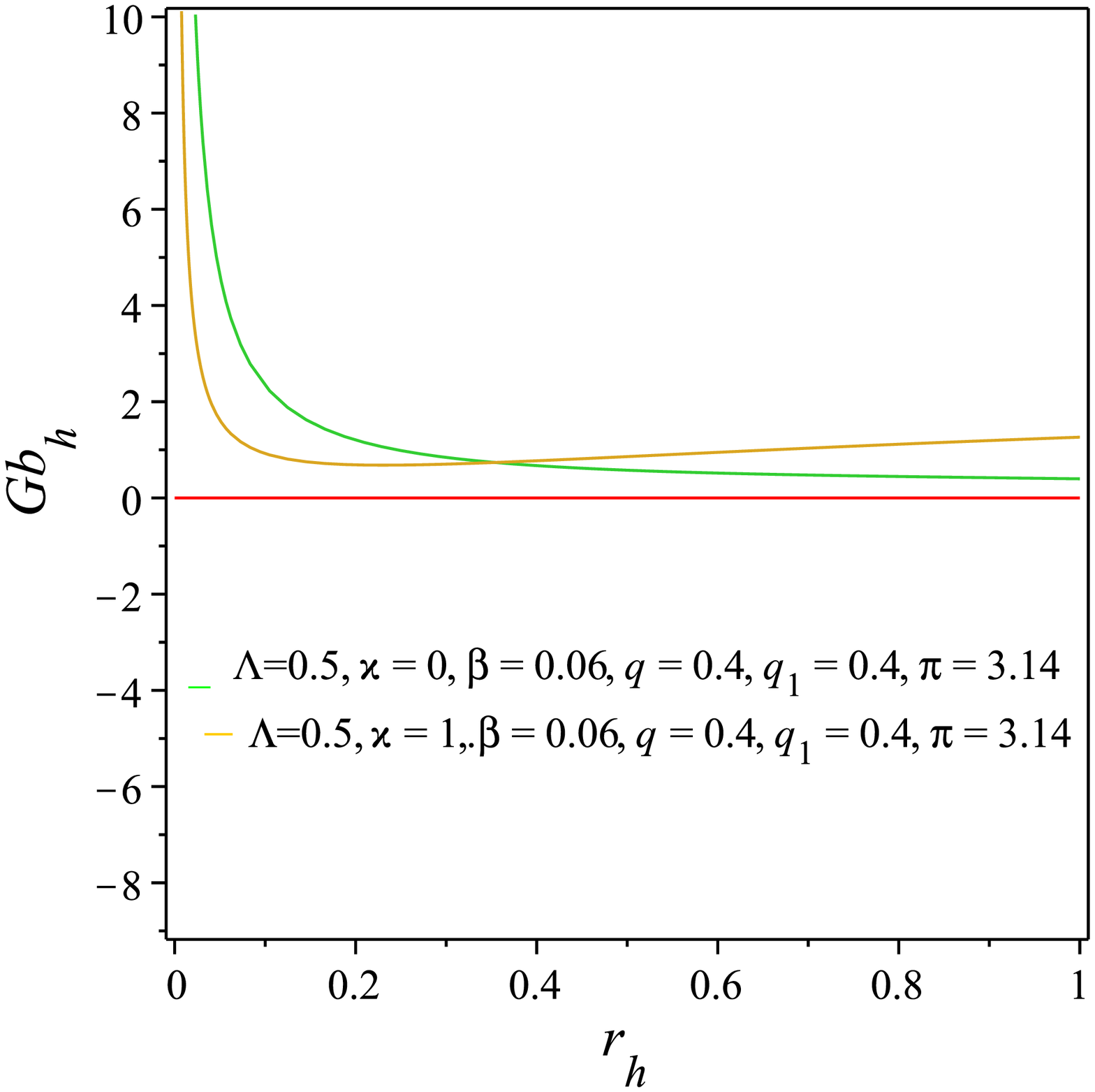}}
\caption{Schematic plots of heat capacity of Eqs.  (\ref{s3}) and (\ref{s44}).}
\label{Fig:5}
\end{figure}
If $\chi \rightarrow 0$, the Gibb's free energy in the non-charged case  becomes equal to that in \cite{Altamirano:2014tva}.
The Gibb's energy for black holes are depicted in Figs. \ref{Fig:5}\subref{fig:5a} and \ref{Fig:5}\subref{fig:5b}. As \ref{Fig:5}\subref{fig:5b} shows that the Gibb's energy is always positive which means that it is  more globally stable than the other three spacetimes.

\section{Summary and discussion}\label{S6}
In this study, we have addressed the  Taub-NUT spacetimes in $f(R)$ gravitational theory. We describe the gravitational field equations for $f(R)=R+\beta R^2$ and apply them to Taub-NUT spacetime, using the fact that the solution of the trace of the gravitational field equations for $f(R)$ gravity gives a constant Ricci scalar, $R=-8\Lambda$.  We have solved the resulting differential equations exactly and show that the solutions of the  neutral Taub-NUT spacetimes did not depend on the dimensional parameter $\beta$. We have repeated the same calculations to the charged field equations of  $f(R)=R+\beta R^2$ and used the same  Taub-NUT spacetime. We have solved the resulting differential equations analytically and show that the output solution  depends on the dimensional parameter $\beta$ and must satisfy the constraint $\beta \neq \frac{1}{16\Lambda}$.

The physical quantities of these black hole solutions are studies. Among different things, we have studied the singularities and show that all the black hole solutions have a singularity at $\varrho=0 \Rightarrow r=0$    when the parameter $\chi=0$.    Also, we have studied the horizons and show that there are two horizons corresponding to the event and the cosmological horizons. We have shown these horizons in  Fig. \ref{Fig:1} for the charged and the non-charged cases.  Furthermore, thermodynamics for black holes has been explored and the thermal phase transition based on the discontinuous sign changing of the specific heat has been investigated. We have calculated the mass in terms of the horizons and have shown the behavior of these quantities in Fig \ref{Fig:2}. Also, we have calculated the temperature in terms of the horizons and indicate their behavior in Fig. \ref{Fig:3}. Moreover, we have calculated the heat capacity of each black hole solution and have  shown their behavior in Fig. \ref{Fig:4}.  We have  shown that the solution of Taub-NUT spacetime is unstable in the region $r<r_h$ and then has a phase transition at $r=r_h$ then has a stable value at $r>r_h$ as  Fig. \ref{Fig:4}\subref{fig:4a} shows. Same discussions can be applied for the other charged Taub-NUT, as Figs. \ref{Fig:4}\subref{fig:4b} shows.  In addition, the free energy for these solutions has been analyzed and the pattern of those has depicted in Fig. \ref{Fig:5}. It has been found from \ref{Fig:5}\subref{fig:5b} that the charged Taub-NUT solution always has local stability \cite{Jawad:2017mwt}.

It is of interest to note that L$\ddot{u}$ et al. \cite{PhysRevD.92.124019} have derived numerical spherically symmetric solution in higher order derivative gravity using the action
\begin{eqnarray} \label{m78} I=\int d^4x \sqrt{-g}[\alpha R-\gamma C_{\mu \nu \rho \beta} C^{\mu \nu \rho \beta}+\beta R^2],\end{eqnarray}
with $\alpha$, $\beta$ and $\gamma$ being constants and $C_{\mu \nu \rho \beta}$ is the Weyl tensor\footnote{In this numerical spherically symmetric solution, $\gamma$-term is significant in deriving solutions.}. It is of interest to note that if $\alpha=1$ and $\gamma=0$ the field equations resulting from L$\ddot{u}$ et al. \cite{PhysRevLett.114.171601} will be identical with the one studied here in the non-charged case. So, is it possible to derive a NUT or Taub-NUT solution in higher order derivative gravity?  Moreover, the stability analysis using geodesic deviation \cite{Nashed:2003ee} of the above black holes needs to be checked. These will be answered elsewhere.

Before we close  this discussion we must stress on the fact that all the black holes derived in this study can easily transformed to Einstein frame using a constant scalar field, due to the fact that
\begin{equation}\label{conf-trans1}
g_{\mu \nu} \to  { g}_{{\mu \nu}_{Ein}}=\Omega^2(x) g_{{\mu \nu}_{Jor}},\quad \textrm where \quad
\Omega^2=f_R,\,
\end{equation}
with $f_R$ being constant value and ${ g}_{{\mu \nu}_{Ein}}$ is the metric in Einstein frame and ${ g}_{{\mu \nu}_{Jor}}$ is the metric in Jordan frame.
\section*{Acknowledgments}
 The work of KB was supported in part by the JSPS KAKENHI Grant Number JP
25800136 and Competitive Research Funds for Fukushima University Faculty
(18RI009).

\centerline{\bf Appendix }\vspace{0.2cm}
In the appendix,  we are going to list the necessary quantities of the spacetimes (\ref{m22}) that are used in the calculations  of the field equations (\ref{fe11}) and (\ref{fec}). The non-vanishing components of the Livi-Civita connection of spacetime (\ref{m22}) are:
\begin{eqnarray} \label{cht}
&&\Big\{^r_{ t t}\Big\}=\frac{N_1s'}{2}, \qquad  \Big\{^t_{ r t}\Big\}=\frac{s'}{2s},\qquad \Big\{^t_{ \theta t}\Big\}=-\frac{2\chi^2 s \cos\theta}{k\sin\theta},  \qquad \Big\{^\phi_{ t \theta}\Big\}=\frac{\chi s }{k\sin\theta}, \qquad \Big\{^r_{ \phi t}\Big\}=\chi N_1s'\cos\theta, \qquad \Big\{^r_{ r r}\Big\}=-\frac{N'_1}{2N_1},  \nonumber\\
&&   \Big\{^\theta_{ t \phi}\Big\}=-\frac{\chi s\sin\theta}{k},\qquad \Big\{^\phi_{ r \phi}\Big\}=\Big\{^\theta_{ r \theta}\Big\}=\frac{k'}{2k}, \qquad \Big\{^t_{ r \phi}\Big\}=\frac{\chi \cos\theta [k's-s'k]}{ks}, \qquad \Big\{^r_{ \theta \theta}\Big\}=-\frac{N_1 k'}{2}, \qquad \Big\{^\phi_{ \phi \theta}\Big\}=\frac{\cos\theta(2\chi^2s+k)}{k\sin\theta},\nonumber\\
&&  \qquad \Big\{^t_{ \phi \theta}\Big\}=-\frac{\chi[k+\cos^2\theta(4\chi^2s+k)]}{k\sin\theta}, \qquad \Big\{^r_{ \phi \phi}\Big\}=\frac{N_1[4\chi^2s'\cos^2\theta-k'\sin^2\theta]}{2}, \qquad \Big\{^\theta_{ \phi \phi}\Big\}=-\frac{\sin\theta\cos\theta(4\chi^2s+k)}{k}.
\end{eqnarray}
The non-zero components of the Riemann tensor for (\ref{m22}) are:
\begin{eqnarray} \label{Rie}
&& R_{trtr}=\frac{2sN_1s''-N_1s'^2+3ss'N'_1+2s^2N''_1}{4s}, \qquad R_{trr\phi}=\frac{\chi \cos\theta(N_1s'^2-2sN_1s''-3ss'N'_1-2s^2N''_1)}{2s},\nonumber\\
&&  R_{tr\theta\phi}= 2R_{t\theta r\phi}=-2R_{t\phi r\theta}=\frac{\chi \sin\theta(kN_1s'+ksN'_1-sN_1k')}{k}, \qquad R_{r\phi \theta \phi}=\frac{3\chi^2 \sin2\theta(sN_1k'-kN_1s'-ksN'_1)}{2k},
\nonumber\\
&& R_{t\theta t\theta}= R_{t\phi t\phi}\sin^2\theta=\frac{R_{t\theta \theta \phi}}{2\chi \cos\theta}=\frac{N_1[kN_1s'k'+ksk'N'_1+4\chi^2s^2N_1]}{4k},\nonumber\\
&&   R_{r\theta r\theta}=\frac{N_1k'^2-2N_1kk''-kk'N'_1}{4N_1k}, \qquad  R_{\theta \phi \theta \phi}=\frac{[12\chi^2 ksN_1+4k^2-N_1kk'^2]\sin^2\theta+4N_1\chi^2\cos^2\theta[4\chi^2s^2N_1+N_1kk's'+skk'N'_1]}{4k},
\nonumber\\
&&   R_{r\phi r\phi}=\frac{4\chi^2k \cos^2\theta[2sN_1{}^2s''+2s^2N_1N''_1-N_1{}^2s'^2+3sN_1s'N'_1]-\sin^2\theta[2sN_1kk''+ksN'_1k'-sN_1k'^2]}{4N_1sk}.
\end{eqnarray}
The non-zero components of the Riemann tensor for (\ref{m22}) are:
\begin{eqnarray} \label{Rie}
&& R_{tt}=\frac{N_1(2k^2sN_1s''-N_1k^2s'^2+3k^2ss'N'_1+2k^2s^2N''_1+2ksN_1k's'+2ks^2k'N'_1+8\chi^2 N_1s^3)}{4sk^2},\nonumber\\
&&  R_{t\phi}=\frac{\chi N_1\cos\theta(3sk^2N'_1s'+2k^2sN_1s''+2k^2sN_1s''-N_1k^2s'^2+2ksN_1s'k'+2ks^2k'N'_1+8\chi^2s^3N_1)}{2sk^2},\nonumber\\
&&  R_{\theta \theta}=\frac{4ks-kN_1 s'k'-2ksk'N'_1-2ksN_1k''+8\chi^2 s^2N_1}{4ks},\nonumber\\
&&  R_{r r}=\frac{2s^2N_1k'^2-4ks^2N_1kk''-2s^2kk'N'_1-2k^2sN_1s''-3k^2ss'N'_1+k^2N_1s'^2-2k^2s^2N''_1}{4s^2N_1k^2},\nonumber\\
&&  R_{\phi \phi}=\frac{1}{4sk^2}\Bigg(4\chi^2  N_1\cos^2\theta  [2k^2sN_1s''+2ks N_1s'k'+2s^2kk'N'_1+8\chi^2 s^3N_1 -k^2 N_1s'^2+3s k^2 s'N'_1+2k^2s^2N''_1]\nonumber\\
&&-\sin^2\theta[2k^2sN_1k''+2k^2 sN'_1k'+N_1k^2 s'k'-8\chi^2 s^2N_1k+4k^2s]\Bigg).
\end{eqnarray}


\begin{thebibliography}{102}%
\makeatletter
\providecommand \@ifxundefined [1]{%
 \@ifx{#1\undefined}
}%
\providecommand \@ifnum [1]{%
 \ifnum #1\expandafter \@firstoftwo
 \else \expandafter \@secondoftwo
 \fi
}%
\providecommand \@ifx [1]{%
 \ifx #1\expandafter \@firstoftwo
 \else \expandafter \@secondoftwo
 \fi
}%
\providecommand \natexlab [1]{#1}%
\providecommand \enquote  [1]{``#1''}%
\providecommand \bibnamefont  [1]{#1}%
\providecommand \bibfnamefont [1]{#1}%
\providecommand \citenamefont [1]{#1}%
\providecommand \href@noop [0]{\@secondoftwo}%
\providecommand \href [0]{\begingroup \@sanitize@url \@href}%
\providecommand \@href[1]{\@@startlink{#1}\@@href}%
\providecommand \@@href[1]{\endgroup#1\@@endlink}%
\providecommand \@sanitize@url [0]{\catcode `\\12\catcode `\$12\catcode
  `\&12\catcode `\#12\catcode `\^12\catcode `\_12\catcode `\%12\relax}%
\providecommand \@@startlink[1]{}%
\providecommand \@@endlink[0]{}%
\providecommand \url  [0]{\begingroup\@sanitize@url \@url }%
\providecommand \@url [1]{\endgroup\@href {#1}{\urlprefix }}%
\providecommand \urlprefix  [0]{URL }%
\providecommand \Eprint [0]{\href }%
\providecommand \doibase [0]{http://dx.doi.org/}%
\providecommand \selectlanguage [0]{\@gobble}%
\providecommand \bibinfo  [0]{\@secondoftwo}%
\providecommand \bibfield  [0]{\@secondoftwo}%
\providecommand \translation [1]{[#1]}%
\providecommand \BibitemOpen [0]{}%
\providecommand \bibitemStop [0]{}%
\providecommand \bibitemNoStop [0]{.\EOS\space}%
\providecommand \EOS [0]{\spacefactor3000\relax}%
\providecommand \BibitemShut  [1]{\csname bibitem#1\endcsname}%
\let\auto@bib@innerbib\@empty
\bibitem [{\citenamefont {Birrell}\ and\ \citenamefont
  {Davies}(1984)}]{Birrell:1982ix}%
  \BibitemOpen
  \bibfield  {author} {\bibinfo {author} {\bibfnamefont {N.~D.}\ \bibnamefont
  {Birrell}}\ and\ \bibinfo {author} {\bibfnamefont {P.~C.~W.}\ \bibnamefont
  {Davies}},\ }\href {\doibase 10.1017/CBO9780511622632} {\emph {\bibinfo
  {title} {{Quantum Fields in Curved Space}}}},\ Cambridge Monographs on
  Mathematical Physics\ (\bibinfo  {publisher} {Cambridge Univ. Press},\
  \bibinfo {address} {Cambridge, UK},\ \bibinfo {year} {1984})\BibitemShut
  {NoStop}%
\bibitem [{\citenamefont {Starobinsky}(1980)}]{Starobinsky:1980te}%
  \BibitemOpen
  \bibfield  {author} {\bibinfo {author} {\bibfnamefont {Alexei~A.}\
  \bibnamefont {Starobinsky}},\ }\bibfield  {title} {\enquote {\bibinfo {title}
  {{A New Type of Isotropic Cosmological Models Without Singularity}},}\ }\href
  {\doibase 10.1016/0370-2693(80)90670-X} {\bibfield  {journal} {\bibinfo
  {journal} {Phys. Lett.}\ }\textbf {\bibinfo {volume} {B91}},\ \bibinfo
  {pages} {99--102} (\bibinfo {year} {1980})},\ \bibinfo {note}
  {[,771(1980)]}\BibitemShut {NoStop}%
\bibitem [{\citenamefont {Psaltis}\ \emph {et~al.}(2008)\citenamefont
  {Psaltis}, \citenamefont {Perrodin}, \citenamefont {Dienes},\ and\
  \citenamefont {Mocioiu}}]{Psaltis:2007cw}%
  \BibitemOpen
  \bibfield  {author} {\bibinfo {author} {\bibfnamefont {Dimitrios}\
  \bibnamefont {Psaltis}}, \bibinfo {author} {\bibfnamefont {Delphine}\
  \bibnamefont {Perrodin}}, \bibinfo {author} {\bibfnamefont {Keith~R.}\
  \bibnamefont {Dienes}}, \ and\ \bibinfo {author} {\bibfnamefont {Irina}\
  \bibnamefont {Mocioiu}},\ }\bibfield  {title} {\enquote {\bibinfo {title}
  {{Kerr Black Holes are Not Unique to General Relativity}},}\ }\href {\doibase
  10.1103/PhysRevLett.100.091101, 10.1103/PhysRevLett.100.119902} {\bibfield
  {journal} {\bibinfo  {journal} {Phys. Rev. Lett.}\ }\textbf {\bibinfo
  {volume} {100}},\ \bibinfo {pages} {091101} (\bibinfo {year} {2008})},\
  \bibinfo {note} {[Phys. Rev. Lett.100,119902(2008)]},\ \Eprint
  {http://arxiv.org/abs/0710.4564} {arXiv:0710.4564 [astro-ph]} \BibitemShut
  {NoStop}%
\bibitem [{\citenamefont {Capozziello}\ \emph {et~al.}(2003)\citenamefont
  {Capozziello}, \citenamefont {Carloni},\ and\ \citenamefont
  {Troisi}}]{Capozziello:2003tk}%
  \BibitemOpen
  \bibfield  {author} {\bibinfo {author} {\bibfnamefont {Salvatore}\
  \bibnamefont {Capozziello}}, \bibinfo {author} {\bibfnamefont {Sante}\
  \bibnamefont {Carloni}}, \ and\ \bibinfo {author} {\bibfnamefont {Antonio}\
  \bibnamefont {Troisi}},\ }\bibfield  {title} {\enquote {\bibinfo {title}
  {{Quintessence without scalar fields}},}\ }\href@noop {} {\bibfield
  {journal} {\bibinfo  {journal} {Recent Res. Dev. Astron. Astrophys.}\
  }\textbf {\bibinfo {volume} {1}},\ \bibinfo {pages} {625} (\bibinfo {year}
  {2003})},\ \Eprint {http://arxiv.org/abs/astro-ph/0303041}
  {arXiv:astro-ph/0303041 [astro-ph]} \BibitemShut {NoStop}%
\bibitem [{\citenamefont {Carroll}\ \emph {et~al.}(2004)\citenamefont
  {Carroll}, \citenamefont {Duvvuri}, \citenamefont {Trodden},\ and\
  \citenamefont {Turner}}]{Carroll:2003wy}%
  \BibitemOpen
  \bibfield  {author} {\bibinfo {author} {\bibfnamefont {Sean~M.}\ \bibnamefont
  {Carroll}}, \bibinfo {author} {\bibfnamefont {Vikram}\ \bibnamefont
  {Duvvuri}}, \bibinfo {author} {\bibfnamefont {Mark}\ \bibnamefont {Trodden}},
  \ and\ \bibinfo {author} {\bibfnamefont {Michael~S.}\ \bibnamefont
  {Turner}},\ }\bibfield  {title} {\enquote {\bibinfo {title} {{Is cosmic speed
  - up due to new gravitational physics?}}}\ }\href {\doibase
  10.1103/PhysRevD.70.043528} {\bibfield  {journal} {\bibinfo  {journal} {Phys.
  Rev.}\ ,\ \bibinfo {pages} {043528}} (\bibinfo {year} {2004})},\ \Eprint
  {http://arxiv.org/abs/astro-ph/0306438} {arXiv:astro-ph/0306438 [astro-ph]}
  \BibitemShut {NoStop}%
\bibitem [{\citenamefont {Nojiri}\ and\ \citenamefont
  {Odintsov}(2003)}]{Nojiri:2003ft}%
  \BibitemOpen
  \bibfield  {author} {\bibinfo {author} {\bibfnamefont {Shin'ichi}\
  \bibnamefont {Nojiri}}\ and\ \bibinfo {author} {\bibfnamefont {Sergei~D.}\
  \bibnamefont {Odintsov}},\ }\bibfield  {title} {\enquote {\bibinfo {title}
  {{Modified gravity with negative and positive powers of the curvature:
  Unification of the inflation and of the cosmic acceleration}},}\ }\href
  {\doibase 10.1103/PhysRevD.68.123512} {\bibfield  {journal} {\bibinfo
  {journal} {Phys. Rev.}\ ,\ \bibinfo {pages} {123512}} (\bibinfo {year}
  {2003})},\ \Eprint {http://arxiv.org/abs/hep-th/0307288}
  {arXiv:hep-th/0307288 [hep-th]} \BibitemShut {NoStop}%
\bibitem [{\citenamefont {De~Felice}\ and\ \citenamefont
  {Tsujikawa}(2010)}]{DeFelice:2010aj}%
  \BibitemOpen
  \bibfield  {author} {\bibinfo {author} {\bibfnamefont {Antonio}\ \bibnamefont
  {De~Felice}}\ and\ \bibinfo {author} {\bibfnamefont {Shinji}\ \bibnamefont
  {Tsujikawa}},\ }\bibfield  {title} {\enquote {\bibinfo {title} {{f(R)
  theories}},}\ }\href {\doibase 10.12942/lrr-2010-3} {\bibfield  {journal}
  {\bibinfo  {journal} {Living Rev. Rel.}\ }\textbf {\bibinfo {volume} {13}},\
  \bibinfo {pages} {3} (\bibinfo {year} {2010})},\ \Eprint
  {http://arxiv.org/abs/1002.4928} {arXiv:1002.4928 [gr-qc]} \BibitemShut
  {NoStop}%
\bibitem [{\citenamefont {{Sotiriou}}\ and\ \citenamefont
  {{Faraoni}}(2010)}]{2010RvMP...82..451S}%
  \BibitemOpen
  \bibfield  {author} {\bibinfo {author} {\bibfnamefont {T.~P.}\ \bibnamefont
  {{Sotiriou}}}\ and\ \bibinfo {author} {\bibfnamefont {V.}~\bibnamefont
  {{Faraoni}}},\ }\bibfield  {title} {\enquote {\bibinfo {title} {{f(R)
  theories of gravity}},}\ }\href {\doibase 10.1103/RevModPhys.82.451}
  {\bibfield  {journal} {\bibinfo  {journal} {Reviews of Modern Physics}\
  }\textbf {\bibinfo {volume} {82}},\ \bibinfo {pages} {451--497} (\bibinfo
  {year} {2010})},\ \Eprint {http://arxiv.org/abs/0805.1726} {arXiv:0805.1726
  [gr-qc]} \BibitemShut {NoStop}%
\bibitem [{\citenamefont {Vignolo}\ \emph {et~al.}(2018)\citenamefont
  {Vignolo}, \citenamefont {Cianci},\ and\ \citenamefont
  {Carloni}}]{Vignolo:2018eco}%
  \BibitemOpen
  \bibfield  {author} {\bibinfo {author} {\bibfnamefont {Stefano}\ \bibnamefont
  {Vignolo}}, \bibinfo {author} {\bibfnamefont {Roberto}\ \bibnamefont
  {Cianci}}, \ and\ \bibinfo {author} {\bibfnamefont {Sante}\ \bibnamefont
  {Carloni}},\ }\bibfield  {title} {\enquote {\bibinfo {title} {{On the
  junction conditions in $f(R)$-gravity with torsion}},}\ }\href {\doibase
  10.1088/1361-6382/aab6fe} {\bibfield  {journal} {\bibinfo  {journal} {Class.
  Quant. Grav.}\ }\textbf {\bibinfo {volume} {35}},\ \bibinfo {pages} {095014}
  (\bibinfo {year} {2018})},\ \Eprint {http://arxiv.org/abs/1801.08344}
  {arXiv:1801.08344 [gr-qc]} \BibitemShut {NoStop}%
\bibitem [{\citenamefont {Capozziello}\ and\ \citenamefont
  {Francaviglia}(2008)}]{Capozziello:2007ec}%
  \BibitemOpen
  \bibfield  {author} {\bibinfo {author} {\bibfnamefont {Salvatore}\
  \bibnamefont {Capozziello}}\ and\ \bibinfo {author} {\bibfnamefont {Mauro}\
  \bibnamefont {Francaviglia}},\ }\bibfield  {title} {\enquote {\bibinfo
  {title} {{Extended Theories of Gravity and their Cosmological and
  Astrophysical Applications}},}\ }\href {\doibase 10.1007/s10714-007-0551-y}
  {\bibfield  {journal} {\bibinfo  {journal} {Gen. Rel. Grav.}\ }\textbf
  {\bibinfo {volume} {40}},\ \bibinfo {pages} {357--420} (\bibinfo {year}
  {2008})},\ \Eprint {http://arxiv.org/abs/0706.1146} {arXiv:0706.1146
  [astro-ph]} \BibitemShut {NoStop}%
\bibitem [{\citenamefont {Nojiri}\ and\ \citenamefont
  {Odintsov}(2011)}]{Nojiri:2010wj}%
  \BibitemOpen
  \bibfield  {author} {\bibinfo {author} {\bibfnamefont {Shin'ichi}\
  \bibnamefont {Nojiri}}\ and\ \bibinfo {author} {\bibfnamefont {Sergei~D.}\
  \bibnamefont {Odintsov}},\ }\bibfield  {title} {\enquote {\bibinfo {title}
  {{Unified cosmic history in modified gravity: from F(R) theory to Lorentz
  non-invariant models}},}\ }\href {\doibase 10.1016/j.physrep.2011.04.001}
  {\bibfield  {journal} {\bibinfo  {journal} {Phys. Rept.}\ }\textbf {\bibinfo
  {volume} {505}},\ \bibinfo {pages} {59--144} (\bibinfo {year} {2011})},\
  \Eprint {http://arxiv.org/abs/1011.0544} {arXiv:1011.0544 [gr-qc]}
  \BibitemShut {NoStop}%
\bibitem [{\citenamefont {Nojiri}\ \emph {et~al.}(2017)\citenamefont {Nojiri},
  \citenamefont {Odintsov},\ and\ \citenamefont {Oikonomou}}]{Nojiri:2017ncd}%
  \BibitemOpen
  \bibfield  {author} {\bibinfo {author} {\bibfnamefont {S.}~\bibnamefont
  {Nojiri}}, \bibinfo {author} {\bibfnamefont {S.~D.}\ \bibnamefont
  {Odintsov}}, \ and\ \bibinfo {author} {\bibfnamefont {V.~K.}\ \bibnamefont
  {Oikonomou}},\ }\bibfield  {title} {\enquote {\bibinfo {title} {{Modified
  Gravity Theories on a Nutshell: Inflation, Bounce and Late-time
  Evolution}},}\ }\href {\doibase 10.1016/j.physrep.2017.06.001} {\bibfield
  {journal} {\bibinfo  {journal} {Phys. Rept.}\ }\textbf {\bibinfo {volume}
  {692}},\ \bibinfo {pages} {1--104} (\bibinfo {year} {2017})},\ \Eprint
  {http://arxiv.org/abs/1705.11098} {arXiv:1705.11098 [gr-qc]} \BibitemShut
  {NoStop}%
\bibitem [{\citenamefont {Capozziello}\ and\ \citenamefont
  {De~Laurentis}(2011)}]{Capozziello:2011et}%
  \BibitemOpen
  \bibfield  {author} {\bibinfo {author} {\bibfnamefont {Salvatore}\
  \bibnamefont {Capozziello}}\ and\ \bibinfo {author} {\bibfnamefont
  {Mariafelicia}\ \bibnamefont {De~Laurentis}},\ }\bibfield  {title} {\enquote
  {\bibinfo {title} {{Extended Theories of Gravity}},}\ }\href {\doibase
  10.1016/j.physrep.2011.09.003} {\bibfield  {journal} {\bibinfo  {journal}
  {Phys. Rept.}\ }\textbf {\bibinfo {volume} {509}},\ \bibinfo {pages}
  {167--321} (\bibinfo {year} {2011})},\ \Eprint
  {http://arxiv.org/abs/1108.6266} {arXiv:1108.6266 [gr-qc]} \BibitemShut
  {NoStop}%
\bibitem [{\citenamefont {Faraoni}\ and\ \citenamefont
  {Capozziello}(2011)}]{Capozziello:2010zz}%
  \BibitemOpen
  \bibfield  {author} {\bibinfo {author} {\bibfnamefont {Valerio}\ \bibnamefont
  {Faraoni}}\ and\ \bibinfo {author} {\bibfnamefont {Salvatore}\ \bibnamefont
  {Capozziello}},\ }\href {\doibase 10.1007/978-94-007-0165-6} {\emph {\bibinfo
  {title} {{Beyond Einstein Gravity}}}},\ Vol.\ \bibinfo {volume} {170}\
  (\bibinfo  {publisher} {Springer},\ \bibinfo {address} {Dordrecht},\ \bibinfo
  {year} {2011})\BibitemShut {NoStop}%
\bibitem [{\citenamefont {Bamba}\ and\ \citenamefont
  {Odintsov}(2015)}]{Bamba:2015uma}%
  \BibitemOpen
  \bibfield  {author} {\bibinfo {author} {\bibfnamefont {Kazuharu}\
  \bibnamefont {Bamba}}\ and\ \bibinfo {author} {\bibfnamefont {Sergei~D.}\
  \bibnamefont {Odintsov}},\ }\bibfield  {title} {\enquote {\bibinfo {title}
  {{Inflationary cosmology in modified gravity theories}},}\ }\href {\doibase
  10.3390/sym7010220} {\bibfield  {journal} {\bibinfo  {journal} {Symmetry}\
  }\textbf {\bibinfo {volume} {7}},\ \bibinfo {pages} {220--240} (\bibinfo
  {year} {2015})},\ \Eprint {http://arxiv.org/abs/1503.00442} {arXiv:1503.00442
  [hep-th]} \BibitemShut {NoStop}%
\bibitem [{\citenamefont {Cai}\ \emph {et~al.}(2016)\citenamefont {Cai},
  \citenamefont {Capozziello}, \citenamefont {De~Laurentis},\ and\
  \citenamefont {Saridakis}}]{Cai:2015emx}%
  \BibitemOpen
  \bibfield  {author} {\bibinfo {author} {\bibfnamefont {Yi-Fu}\ \bibnamefont
  {Cai}}, \bibinfo {author} {\bibfnamefont {Salvatore}\ \bibnamefont
  {Capozziello}}, \bibinfo {author} {\bibfnamefont {Mariafelicia}\ \bibnamefont
  {De~Laurentis}}, \ and\ \bibinfo {author} {\bibfnamefont {Emmanuel~N.}\
  \bibnamefont {Saridakis}},\ }\bibfield  {title} {\enquote {\bibinfo {title}
  {{f(T) teleparallel gravity and cosmology}},}\ }\href {\doibase
  10.1088/0034-4885/79/10/106901} {\bibfield  {journal} {\bibinfo  {journal}
  {Rept. Prog. Phys.}\ }\textbf {\bibinfo {volume} {79}},\ \bibinfo {pages}
  {106901} (\bibinfo {year} {2016})},\ \Eprint
  {http://arxiv.org/abs/1511.07586} {arXiv:1511.07586 [gr-qc]} \BibitemShut
  {NoStop}%
\bibitem [{\citenamefont {Bamba}\ \emph {et~al.}(2012)\citenamefont {Bamba},
  \citenamefont {Capozziello}, \citenamefont {Nojiri},\ and\ \citenamefont
  {Odintsov}}]{Bamba:2012cp}%
  \BibitemOpen
  \bibfield  {author} {\bibinfo {author} {\bibfnamefont {Kazuharu}\
  \bibnamefont {Bamba}}, \bibinfo {author} {\bibfnamefont {Salvatore}\
  \bibnamefont {Capozziello}}, \bibinfo {author} {\bibfnamefont {Shin'ichi}\
  \bibnamefont {Nojiri}}, \ and\ \bibinfo {author} {\bibfnamefont {Sergei~D.}\
  \bibnamefont {Odintsov}},\ }\bibfield  {title} {\enquote {\bibinfo {title}
  {{Dark energy cosmology: the equivalent description via different theoretical
  models and cosmography tests}},}\ }\href {\doibase 10.1007/s10509-012-1181-8}
  {\bibfield  {journal} {\bibinfo  {journal} {Astrophys. Space Sci.}\ }\textbf
  {\bibinfo {volume} {342}},\ \bibinfo {pages} {155--228} (\bibinfo {year}
  {2012})},\ \Eprint {http://arxiv.org/abs/1205.3421} {arXiv:1205.3421 [gr-qc]}
  \BibitemShut {NoStop}%
\bibitem [{\citenamefont {Akrami}\ \emph {et~al.}(2018)\citenamefont {Akrami}
  \emph {et~al.}}]{Akrami:2018odb}%
  \BibitemOpen
  \bibfield  {author} {\bibinfo {author} {\bibfnamefont {Y.}~\bibnamefont
  {Akrami}} \emph {et~al.} (\bibinfo {collaboration} {Planck}),\ }\bibfield
  {title} {\enquote {\bibinfo {title} {{Planck 2018 results. X. Constraints on
  inflation}},}\ }\href@noop {} {\  (\bibinfo {year} {2018})},\ \Eprint
  {http://arxiv.org/abs/1807.06211} {arXiv:1807.06211 [astro-ph.CO]}
  \BibitemShut {NoStop}%
\bibitem [{\citenamefont {Amendola}\ \emph {et~al.}(2007)\citenamefont
  {Amendola}, \citenamefont {Polarski},\ and\ \citenamefont
  {Tsujikawa}}]{Amendola:2006eh}%
  \BibitemOpen
  \bibfield  {author} {\bibinfo {author} {\bibfnamefont {Luca}\ \bibnamefont
  {Amendola}}, \bibinfo {author} {\bibfnamefont {David}\ \bibnamefont
  {Polarski}}, \ and\ \bibinfo {author} {\bibfnamefont {Shinji}\ \bibnamefont
  {Tsujikawa}},\ }\bibfield  {title} {\enquote {\bibinfo {title} {{Power-laws
  f(R) theories are cosmologically unacceptable}},}\ }\href {\doibase
  10.1142/S0218271807010936} {\bibfield  {journal} {\bibinfo  {journal} {Int.
  J. Mod. Phys.}\ }\textbf {\bibinfo {volume} {D16}},\ \bibinfo {pages}
  {1555--1561} (\bibinfo {year} {2007})},\ \Eprint
  {http://arxiv.org/abs/astro-ph/0605384} {arXiv:astro-ph/0605384 [astro-ph]}
  \BibitemShut {NoStop}%
\bibitem [{\citenamefont {Maselli}\ \emph {et~al.}(2015)\citenamefont
  {Maselli}, \citenamefont {Pani}, \citenamefont {Gualtieri},\ and\
  \citenamefont {Ferrari}}]{PhysRevD.92.083014}%
  \BibitemOpen
  \bibfield  {author} {\bibinfo {author} {\bibfnamefont {Andrea}\ \bibnamefont
  {Maselli}}, \bibinfo {author} {\bibfnamefont {Paolo}\ \bibnamefont {Pani}},
  \bibinfo {author} {\bibfnamefont {Leonardo}\ \bibnamefont {Gualtieri}}, \
  and\ \bibinfo {author} {\bibfnamefont {Valeria}\ \bibnamefont {Ferrari}},\
  }\bibfield  {title} {\enquote {\bibinfo {title} {Rotating black holes in
  einstein-dilaton-gauss-bonnet gravity with finite coupling},}\ }\href
  {\doibase 10.1103/PhysRevD.92.083014} {\bibfield  {journal} {\bibinfo
  {journal} {Phys. Rev. D}\ }\textbf {\bibinfo {volume} {92}},\ \bibinfo
  {pages} {083014} (\bibinfo {year} {2015})}\BibitemShut {NoStop}%
\bibitem [{\citenamefont {Bambi}(2013)}]{Bambi:2013nla}%
  \BibitemOpen
  \bibfield  {author} {\bibinfo {author} {\bibfnamefont {Cosimo}\ \bibnamefont
  {Bambi}},\ }\bibfield  {title} {\enquote {\bibinfo {title} {{Can the
  supermassive objects at the centers of galaxies be traversable wormholes? The
  first test of strong gravity for mm/sub-mm very long baseline interferometry
  facilities}},}\ }\href {\doibase 10.1103/PhysRevD.87.107501} {\bibfield
  {journal} {\bibinfo  {journal} {Phys. Rev.}\ }\textbf {\bibinfo {volume}
  {D87}},\ \bibinfo {pages} {107501} (\bibinfo {year} {2013})},\ \Eprint
  {http://arxiv.org/abs/1304.5691} {arXiv:1304.5691 [gr-qc]} \BibitemShut
  {NoStop}%
\bibitem [{\citenamefont {Yunes}\ and\ \citenamefont
  {Stein}(2011)}]{PhysRevD.83.104002}%
  \BibitemOpen
  \bibfield  {author} {\bibinfo {author} {\bibfnamefont {Nicol\'as}\
  \bibnamefont {Yunes}}\ and\ \bibinfo {author} {\bibfnamefont {Leo~C.}\
  \bibnamefont {Stein}},\ }\bibfield  {title} {\enquote {\bibinfo {title}
  {Nonspinning black holes in alternative theories of gravity},}\ }\href
  {\doibase 10.1103/PhysRevD.83.104002} {\bibfield  {journal} {\bibinfo
  {journal} {Phys. Rev. D}\ }\textbf {\bibinfo {volume} {83}},\ \bibinfo
  {pages} {104002} (\bibinfo {year} {2011})}\BibitemShut {NoStop}%
\bibitem [{\citenamefont {Awad}\ \emph {et~al.}(2017)\citenamefont {Awad},
  \citenamefont {Capozziello},\ and\ \citenamefont {Nashed}}]{Awad:2017tyz}%
  \BibitemOpen
  \bibfield  {author} {\bibinfo {author} {\bibfnamefont {A.~M.}\ \bibnamefont
  {Awad}}, \bibinfo {author} {\bibfnamefont {S.}~\bibnamefont {Capozziello}}, \
  and\ \bibinfo {author} {\bibfnamefont {G.~G.~L.}\ \bibnamefont {Nashed}},\
  }\bibfield  {title} {\enquote {\bibinfo {title} {{$D$-dimensional charged
  Anti-de-Sitter black holes in $f(T)$ gravity}},}\ }\href {\doibase
  10.1007/JHEP07(2017)136} {\bibfield  {journal} {\bibinfo  {journal} {JHEP}\
  }\textbf {\bibinfo {volume} {07}},\ \bibinfo {pages} {136} (\bibinfo {year}
  {2017})},\ \Eprint {http://arxiv.org/abs/1706.01773} {arXiv:1706.01773
  [gr-qc]} \BibitemShut {NoStop}%
\bibitem [{\citenamefont {{Ca{\~n}ate}}(2018)}]{2018CQGra..35b5018C}%
  \BibitemOpen
  \bibfield  {author} {\bibinfo {author} {\bibfnamefont {P.}~\bibnamefont
  {{Ca{\~n}ate}}},\ }\bibfield  {title} {\enquote {\bibinfo {title} {{A no-hair
  theorem for black holes in f(R) gravity}},}\ }\href {\doibase
  10.1088/1361-6382/aa8e2e} {\bibfield  {journal} {\bibinfo  {journal}
  {Classical and Quantum Gravity}\ }\textbf {\bibinfo {volume} {35}},\ \bibinfo
  {eid} {025018} (\bibinfo {year} {2018})}\BibitemShut {NoStop}%
\bibitem [{\citenamefont {Will}\ and\ \citenamefont
  {Yunes}(2004)}]{Will:2004xi}%
  \BibitemOpen
  \bibfield  {author} {\bibinfo {author} {\bibfnamefont {Clifford~M.}\
  \bibnamefont {Will}}\ and\ \bibinfo {author} {\bibfnamefont {Nicolas}\
  \bibnamefont {Yunes}},\ }\bibfield  {title} {\enquote {\bibinfo {title}
  {{Testing alternative theories of gravity using LISA}},}\ }\href {\doibase
  10.1088/0264-9381/21/18/006} {\bibfield  {journal} {\bibinfo  {journal}
  {Class. Quant. Grav.}\ }\textbf {\bibinfo {volume} {21}},\ \bibinfo {pages}
  {4367} (\bibinfo {year} {2004})},\ \Eprint
  {http://arxiv.org/abs/gr-qc/0403100} {arXiv:gr-qc/0403100 [gr-qc]}
  \BibitemShut {NoStop}%
\bibitem [{\citenamefont {Motohashi}\ and\ \citenamefont
  {Minamitsuji}(2018)}]{Motohashi:2018wdq}%
  \BibitemOpen
  \bibfield  {author} {\bibinfo {author} {\bibfnamefont {Hayato}\ \bibnamefont
  {Motohashi}}\ and\ \bibinfo {author} {\bibfnamefont {Masato}\ \bibnamefont
  {Minamitsuji}},\ }\bibfield  {title} {\enquote {\bibinfo {title} {{General
  Relativity solutions in modified gravity}},}\ }\href {\doibase
  10.1016/j.physletb.2018.04.041} {\bibfield  {journal} {\bibinfo  {journal}
  {Phys. Lett.}\ }\textbf {\bibinfo {volume} {B781}},\ \bibinfo {pages}
  {728--734} (\bibinfo {year} {2018})},\ \Eprint
  {http://arxiv.org/abs/1804.01731} {arXiv:1804.01731 [gr-qc]} \BibitemShut
  {NoStop}%
\bibitem [{\citenamefont {Cognola}\ \emph {et~al.}(2005)\citenamefont
  {Cognola}, \citenamefont {Elizalde}, \citenamefont {Nojiri}, \citenamefont
  {Odintsov},\ and\ \citenamefont {Zerbini}}]{Cognola:2005de}%
  \BibitemOpen
  \bibfield  {author} {\bibinfo {author} {\bibfnamefont {Guido}\ \bibnamefont
  {Cognola}}, \bibinfo {author} {\bibfnamefont {Emilio}\ \bibnamefont
  {Elizalde}}, \bibinfo {author} {\bibfnamefont {Shin'ichi}\ \bibnamefont
  {Nojiri}}, \bibinfo {author} {\bibfnamefont {Sergei~D.}\ \bibnamefont
  {Odintsov}}, \ and\ \bibinfo {author} {\bibfnamefont {Sergio}\ \bibnamefont
  {Zerbini}},\ }\bibfield  {title} {\enquote {\bibinfo {title} {{One-loop f(R)
  gravity in de Sitter universe}},}\ }\href {\doibase
  10.1088/1475-7516/2005/02/010} {\bibfield  {journal} {\bibinfo  {journal}
  {JCAP}\ }\textbf {\bibinfo {volume} {0502}},\ \bibinfo {pages} {010}
  (\bibinfo {year} {2005})},\ \Eprint {http://arxiv.org/abs/hep-th/0501096}
  {arXiv:hep-th/0501096 [hep-th]} \BibitemShut {NoStop}%
\bibitem [{\citenamefont {Nojiri}\ and\ \citenamefont
  {Odintsov}(2013)}]{Nojiri:2013su}%
  \BibitemOpen
  \bibfield  {author} {\bibinfo {author} {\bibfnamefont {Shin'ichi}\
  \bibnamefont {Nojiri}}\ and\ \bibinfo {author} {\bibfnamefont {Sergei~D.}\
  \bibnamefont {Odintsov}},\ }\bibfield  {title} {\enquote {\bibinfo {title}
  {{Anti-Evaporation of Schwarzschild-de Sitter Black Holes in $F(R)$
  gravity}},}\ }\href {\doibase 10.1088/0264-9381/30/12/125003} {\bibfield
  {journal} {\bibinfo  {journal} {Class. Quant. Grav.}\ }\textbf {\bibinfo
  {volume} {30}},\ \bibinfo {pages} {125003} (\bibinfo {year} {2013})},\
  \Eprint {http://arxiv.org/abs/1301.2775} {arXiv:1301.2775 [hep-th]}
  \BibitemShut {NoStop}%
\bibitem [{\citenamefont {Nojiri}\ and\ \citenamefont
  {Odintsov}(2014)}]{Nojiri:2014jqa}%
  \BibitemOpen
  \bibfield  {author} {\bibinfo {author} {\bibfnamefont {Shin'ichi}\
  \bibnamefont {Nojiri}}\ and\ \bibinfo {author} {\bibfnamefont {Sergei~D.}\
  \bibnamefont {Odintsov}},\ }\bibfield  {title} {\enquote {\bibinfo {title}
  {{Instabilities and anti-evaporation of Reissner–Nordström black holes in
  modified $F(R)$ gravity}},}\ }\href {\doibase 10.1016/j.physletb.2014.06.070}
  {\bibfield  {journal} {\bibinfo  {journal} {Phys. Lett.}\ }\textbf {\bibinfo
  {volume} {B735}},\ \bibinfo {pages} {376--382} (\bibinfo {year} {2014})},\
  \Eprint {http://arxiv.org/abs/1405.2439} {arXiv:1405.2439 [gr-qc]}
  \BibitemShut {NoStop}%
\bibitem [{\citenamefont {de~la Cruz-Dombriz}\ \emph
  {et~al.}(2009)\citenamefont {de~la Cruz-Dombriz}, \citenamefont {Dobado},\
  and\ \citenamefont {Maroto}}]{PhysRevD.80.124011}%
  \BibitemOpen
  \bibfield  {author} {\bibinfo {author} {\bibfnamefont {A.}~\bibnamefont
  {de~la Cruz-Dombriz}}, \bibinfo {author} {\bibfnamefont {A.}~\bibnamefont
  {Dobado}}, \ and\ \bibinfo {author} {\bibfnamefont {A.~L.}\ \bibnamefont
  {Maroto}},\ }\bibfield  {title} {\enquote {\bibinfo {title} {Black holes in
  $f(r)$ theories},}\ }\href {\doibase 10.1103/PhysRevD.80.124011} {\bibfield
  {journal} {\bibinfo  {journal} {Phys. Rev. D}\ }\textbf {\bibinfo {volume}
  {80}},\ \bibinfo {pages} {124011} (\bibinfo {year} {2009})}\BibitemShut
  {NoStop}%
\bibitem [{\citenamefont {Sheykhi}(2012)}]{PhysRevD.86.024013}%
  \BibitemOpen
  \bibfield  {author} {\bibinfo {author} {\bibfnamefont {Ahmad}\ \bibnamefont
  {Sheykhi}},\ }\bibfield  {title} {\enquote {\bibinfo {title}
  {Higher-dimensional charged $f(r)$ black holes},}\ }\href {\doibase
  10.1103/PhysRevD.86.024013} {\bibfield  {journal} {\bibinfo  {journal} {Phys.
  Rev. D}\ }\textbf {\bibinfo {volume} {86}},\ \bibinfo {pages} {024013}
  (\bibinfo {year} {2012})}\BibitemShut {NoStop}%
\bibitem [{\citenamefont {{Sheykhi}}\ \emph {et~al.}(2013)\citenamefont
  {{Sheykhi}}, \citenamefont {{Salarpour}},\ and\ \citenamefont
  {{Bahrampour}}}]{2013PhyS...87d5004S}%
  \BibitemOpen
  \bibfield  {author} {\bibinfo {author} {\bibfnamefont {A.}~\bibnamefont
  {{Sheykhi}}}, \bibinfo {author} {\bibfnamefont {S.}~\bibnamefont
  {{Salarpour}}}, \ and\ \bibinfo {author} {\bibfnamefont {Y.}~\bibnamefont
  {{Bahrampour}}},\ }\bibfield  {title} {\enquote {\bibinfo {title} {{Rotating
  black strings in f(R)-Maxwell theory}},}\ }\href {\doibase
  10.1088/0031-8949/87/04/045004} {\bibfield  {journal} {\bibinfo  {journal}
  {physscr}\ }\textbf {\bibinfo {volume} {87}},\ \bibinfo {eid} {045004}
  (\bibinfo {year} {2013})},\ \Eprint {http://arxiv.org/abs/1304.3057}
  {arXiv:1304.3057 [physics.gen-ph]} \BibitemShut {NoStop}%
\bibitem [{\citenamefont {{Nojiri}}\ and\ \citenamefont
  {{Odintsov}}(2013)}]{2013CQGra..30l5003N}%
  \BibitemOpen
  \bibfield  {author} {\bibinfo {author} {\bibfnamefont {S.}~\bibnamefont
  {{Nojiri}}}\ and\ \bibinfo {author} {\bibfnamefont {S.~D.}\ \bibnamefont
  {{Odintsov}}},\ }\bibfield  {title} {\enquote {\bibinfo {title}
  {{Anti-evaporation of Schwarzschild-de Sitter black holes in F(R)
  gravity}},}\ }\href {\doibase 10.1088/0264-9381/30/12/125003} {\bibfield
  {journal} {\bibinfo  {journal} {Classical and Quantum Gravity}\ }\textbf
  {\bibinfo {volume} {30}},\ \bibinfo {eid} {125003} (\bibinfo {year}
  {2013})},\ \Eprint {http://arxiv.org/abs/1301.2775} {arXiv:1301.2775
  [hep-th]} \BibitemShut {NoStop}%
\bibitem [{\citenamefont {Nashed}(2018)}]{Nashed:2018piz}%
  \BibitemOpen
  \bibfield  {author} {\bibinfo {author} {\bibfnamefont {G.~G.~L.}\
  \bibnamefont {Nashed}},\ }\bibfield  {title} {\enquote {\bibinfo {title}
  {{Higher Dimensional Charged Black Hole Solutions in $f(R)$ Gravitational
  Theories}},}\ }\href {\doibase 10.1155/2018/7323574} {\bibfield  {journal}
  {\bibinfo  {journal} {Adv. High Energy Phys.}\ }\textbf {\bibinfo {volume}
  {2018}},\ \bibinfo {pages} {7323574} (\bibinfo {year} {2018})}\BibitemShut
  {NoStop}%
\bibitem [{\citenamefont {{Nashed}}(2018{\natexlab{a}})}]{2018EPJP..133...18N}%
  \BibitemOpen
  \bibfield  {author} {\bibinfo {author} {\bibfnamefont {G.~G.~L.}\
  \bibnamefont {{Nashed}}},\ }\bibfield  {title} {\enquote {\bibinfo {title}
  {{Spherically symmetric charged black holes in f(R) gravitational
  theories}},}\ }\href {\doibase 10.1140/epjp/i2018-11849-7} {\bibfield
  {journal} {\bibinfo  {journal} {European Physical Journal Plus}\ }\textbf
  {\bibinfo {volume} {133}},\ \bibinfo {eid} {18} (\bibinfo {year}
  {2018}{\natexlab{a}})}\BibitemShut {NoStop}%
\bibitem [{\citenamefont {{Nashed}}(2018{\natexlab{b}})}]{2018IJMPD..2750074N}%
  \BibitemOpen
  \bibfield  {author} {\bibinfo {author} {\bibfnamefont {G.~G.~L.}\
  \bibnamefont {{Nashed}}},\ }\bibfield  {title} {\enquote {\bibinfo {title}
  {{Rotating charged black hole spacetimes in quadratic f(R) gravitational
  theories}},}\ }\href {\doibase 10.1142/S0218271818500748} {\bibfield
  {journal} {\bibinfo  {journal} {International Journal of Modern Physics D}\
  }\textbf {\bibinfo {volume} {27}},\ \bibinfo {eid} {1850074} (\bibinfo {year}
  {2018}{\natexlab{b}})}\BibitemShut {NoStop}%
\bibitem [{\citenamefont {Nashed}\ and\ \citenamefont
  {Capozziello}(2019)}]{Nashed:2019tuk}%
  \BibitemOpen
  \bibfield  {author} {\bibinfo {author} {\bibfnamefont {Gamal G.~L.}\
  \bibnamefont {Nashed}}\ and\ \bibinfo {author} {\bibfnamefont {Salvatore}\
  \bibnamefont {Capozziello}},\ }\bibfield  {title} {\enquote {\bibinfo {title}
  {{Charged spherically symmetric black holes in $f(R)$ gravity and their
  stability analysis}},}\ }\href {\doibase 10.1103/PhysRevD.99.104018}
  {\bibfield  {journal} {\bibinfo  {journal} {Phys. Rev.}\ }\textbf {\bibinfo
  {volume} {D99}},\ \bibinfo {pages} {104018} (\bibinfo {year} {2019})},\
  \Eprint {http://arxiv.org/abs/1902.06783} {arXiv:1902.06783 [gr-qc]}
  \BibitemShut {NoStop}%
\bibitem [{\citenamefont {Sotiriou}(2006)}]{Sotiriou:2005xe}%
  \BibitemOpen
  \bibfield  {author} {\bibinfo {author} {\bibfnamefont {Thomas~P.}\
  \bibnamefont {Sotiriou}},\ }\bibfield  {title} {\enquote {\bibinfo {title}
  {{The Nearly Newtonian regime in non-linear theories of gravity}},}\ }\href
  {\doibase 10.1007/s10714-006-0328-8} {\bibfield  {journal} {\bibinfo
  {journal} {Gen. Rel. Grav.}\ }\textbf {\bibinfo {volume} {38}},\ \bibinfo
  {pages} {1407--1417} (\bibinfo {year} {2006})},\ \Eprint
  {http://arxiv.org/abs/gr-qc/0507027} {arXiv:gr-qc/0507027 [gr-qc]}
  \BibitemShut {NoStop}%
\bibitem [{\citenamefont {Faraoni}(2006)}]{PhysRevD.74.023529}%
  \BibitemOpen
  \bibfield  {author} {\bibinfo {author} {\bibfnamefont {Valerio}\ \bibnamefont
  {Faraoni}},\ }\bibfield  {title} {\enquote {\bibinfo {title} {Solar system
  experiments do not yet veto modified gravity models},}\ }\href {\doibase
  10.1103/PhysRevD.74.023529} {\bibfield  {journal} {\bibinfo  {journal} {Phys.
  Rev. D}\ }\textbf {\bibinfo {volume} {74}},\ \bibinfo {pages} {023529}
  (\bibinfo {year} {2006})}\BibitemShut {NoStop}%
\bibitem [{\citenamefont {Nojiri}\ and\ \citenamefont
  {Odintsov}(2006)}]{Nojiri:2006gh}%
  \BibitemOpen
  \bibfield  {author} {\bibinfo {author} {\bibfnamefont {Shin'ichi}\
  \bibnamefont {Nojiri}}\ and\ \bibinfo {author} {\bibfnamefont {Sergei~D.}\
  \bibnamefont {Odintsov}},\ }\bibfield  {title} {\enquote {\bibinfo {title}
  {{Modified f(R) gravity consistent with realistic cosmology: From matter
  dominated epoch to dark energy universe}},}\ }\href {\doibase
  10.1103/PhysRevD.74.086005} {\bibfield  {journal} {\bibinfo  {journal} {Phys.
  Rev.}\ }\textbf {\bibinfo {volume} {D74}},\ \bibinfo {pages} {086005}
  (\bibinfo {year} {2006})},\ \Eprint {http://arxiv.org/abs/hep-th/0608008}
  {arXiv:hep-th/0608008 [hep-th]} \BibitemShut {NoStop}%
\bibitem [{\citenamefont {Sawicki}\ and\ \citenamefont
  {Hu}(2007)}]{Sawicki:2007tf}%
  \BibitemOpen
  \bibfield  {author} {\bibinfo {author} {\bibfnamefont {Ignacy}\ \bibnamefont
  {Sawicki}}\ and\ \bibinfo {author} {\bibfnamefont {Wayne}\ \bibnamefont
  {Hu}},\ }\bibfield  {title} {\enquote {\bibinfo {title} {{Stability of
  Cosmological Solution in f(R) Models of Gravity}},}\ }\href {\doibase
  10.1103/PhysRevD.75.127502} {\bibfield  {journal} {\bibinfo  {journal} {Phys.
  Rev.}\ }\textbf {\bibinfo {volume} {D75}},\ \bibinfo {pages} {127502}
  (\bibinfo {year} {2007})},\ \Eprint {http://arxiv.org/abs/astro-ph/0702278}
  {arXiv:astro-ph/0702278 [astro-ph]} \BibitemShut {NoStop}%
\bibitem [{\citenamefont {de~la Cruz-Dombriz}\ and\ \citenamefont
  {Dobado}(2006)}]{PhysRevD.74.087501}%
  \BibitemOpen
  \bibfield  {author} {\bibinfo {author} {\bibfnamefont {\'Alvaro}\
  \bibnamefont {de~la Cruz-Dombriz}}\ and\ \bibinfo {author} {\bibfnamefont
  {Antonio}\ \bibnamefont {Dobado}},\ }\bibfield  {title} {\enquote {\bibinfo
  {title} {$f(r)$ gravity without a cosmological constant},}\ }\href {\doibase
  10.1103/PhysRevD.74.087501} {\bibfield  {journal} {\bibinfo  {journal} {Phys.
  Rev. D}\ }\textbf {\bibinfo {volume} {74}},\ \bibinfo {pages} {087501}
  (\bibinfo {year} {2006})}\BibitemShut {NoStop}%
\bibitem [{\citenamefont {Dunsby}\ \emph {et~al.}(2010)\citenamefont {Dunsby},
  \citenamefont {Elizalde}, \citenamefont {Goswami}, \citenamefont {Odintsov},\
  and\ \citenamefont {Saez-Gomez}}]{PhysRevD.82.023519}%
  \BibitemOpen
  \bibfield  {author} {\bibinfo {author} {\bibfnamefont {Peter K.~S.}\
  \bibnamefont {Dunsby}}, \bibinfo {author} {\bibfnamefont {Emilio}\
  \bibnamefont {Elizalde}}, \bibinfo {author} {\bibfnamefont {Rituparno}\
  \bibnamefont {Goswami}}, \bibinfo {author} {\bibfnamefont {Sergei}\
  \bibnamefont {Odintsov}}, \ and\ \bibinfo {author} {\bibfnamefont {Diego}\
  \bibnamefont {Saez-Gomez}},\ }\bibfield  {title} {\enquote {\bibinfo {title}
  {$\ensuremath{\Lambda}\mathrm{CDM}$ universe in $f(r)$ gravity},}\ }\href
  {\doibase 10.1103/PhysRevD.82.023519} {\bibfield  {journal} {\bibinfo
  {journal} {Phys. Rev. D}\ }\textbf {\bibinfo {volume} {82}},\ \bibinfo
  {pages} {023519} (\bibinfo {year} {2010})}\BibitemShut {NoStop}%
\bibitem [{\citenamefont {Sebastiani}\ and\ \citenamefont
  {Zerbini}(2011)}]{Sebastiani:2010kv}%
  \BibitemOpen
  \bibfield  {author} {\bibinfo {author} {\bibfnamefont {Lorenzo}\ \bibnamefont
  {Sebastiani}}\ and\ \bibinfo {author} {\bibfnamefont {Sergio}\ \bibnamefont
  {Zerbini}},\ }\bibfield  {title} {\enquote {\bibinfo {title} {{Static
  Spherically Symmetric Solutions in F(R) Gravity}},}\ }\href {\doibase
  10.1140/epjc/s10052-011-1591-8} {\bibfield  {journal} {\bibinfo  {journal}
  {Eur. Phys. J.}\ }\textbf {\bibinfo {volume} {C71}},\ \bibinfo {pages} {1591}
  (\bibinfo {year} {2011})},\ \Eprint {http://arxiv.org/abs/1012.5230}
  {arXiv:1012.5230 [gr-qc]} \BibitemShut {NoStop}%
\bibitem [{\citenamefont {Clifton}\ and\ \citenamefont
  {Barrow}(2005)}]{PhysRevD.72.103005}%
  \BibitemOpen
  \bibfield  {author} {\bibinfo {author} {\bibfnamefont {Timothy}\ \bibnamefont
  {Clifton}}\ and\ \bibinfo {author} {\bibfnamefont {John~D.}\ \bibnamefont
  {Barrow}},\ }\bibfield  {title} {\enquote {\bibinfo {title} {The power of
  general relativity},}\ }\href {\doibase 10.1103/PhysRevD.72.103005}
  {\bibfield  {journal} {\bibinfo  {journal} {Phys. Rev. D}\ }\textbf {\bibinfo
  {volume} {72}},\ \bibinfo {pages} {103005} (\bibinfo {year}
  {2005})}\BibitemShut {NoStop}%
\bibitem [{\citenamefont {Shojai}\ and\ \citenamefont
  {Shojai}(2012)}]{Shojai:2011yq}%
  \BibitemOpen
  \bibfield  {author} {\bibinfo {author} {\bibfnamefont {Ali}\ \bibnamefont
  {Shojai}}\ and\ \bibinfo {author} {\bibfnamefont {Fatimah}\ \bibnamefont
  {Shojai}},\ }\bibfield  {title} {\enquote {\bibinfo {title} {{Some static
  spherically symmetric interior solutions of $f(R)$ gravity}},}\ }\href
  {\doibase 10.1007/s10714-011-1271-x} {\bibfield  {journal} {\bibinfo
  {journal} {Gen. Rel. Grav.}\ }\textbf {\bibinfo {volume} {44}},\ \bibinfo
  {pages} {211--223} (\bibinfo {year} {2012})},\ \Eprint
  {http://arxiv.org/abs/1109.2190} {arXiv:1109.2190 [gr-qc]} \BibitemShut
  {NoStop}%
\bibitem [{\citenamefont {Capozziello}\ \emph {et~al.}(2012)\citenamefont
  {Capozziello}, \citenamefont {Frusciante},\ and\ \citenamefont
  {Vernieri}}]{Capozziello2012}%
  \BibitemOpen
  \bibfield  {author} {\bibinfo {author} {\bibfnamefont {Salvatore}\
  \bibnamefont {Capozziello}}, \bibinfo {author} {\bibfnamefont {Noemi}\
  \bibnamefont {Frusciante}}, \ and\ \bibinfo {author} {\bibfnamefont
  {Daniele}\ \bibnamefont {Vernieri}},\ }\bibfield  {title} {\enquote {\bibinfo
  {title} {New spherically symmetric solutions in f (r)-gravity by noether
  symmetries},}\ }\href {\doibase 10.1007/s10714-012-1367-y} {\bibfield
  {journal} {\bibinfo  {journal} {General Relativity and Gravitation}\ }\textbf
  {\bibinfo {volume} {44}},\ \bibinfo {pages} {1881--1891} (\bibinfo {year}
  {2012})}\BibitemShut {NoStop}%
\bibitem [{\citenamefont {Multam\"aki}\ and\ \citenamefont
  {Vilja}(2006)}]{PhysRevD.74.064022}%
  \BibitemOpen
  \bibfield  {author} {\bibinfo {author} {\bibfnamefont {T.}~\bibnamefont
  {Multam\"aki}}\ and\ \bibinfo {author} {\bibfnamefont {I.}~\bibnamefont
  {Vilja}},\ }\bibfield  {title} {\enquote {\bibinfo {title} {Spherically
  symmetric solutions of modified field equations in $f(r)$ theories of
  gravity},}\ }\href {\doibase 10.1103/PhysRevD.74.064022} {\bibfield
  {journal} {\bibinfo  {journal} {Phys. Rev. D}\ }\textbf {\bibinfo {volume}
  {74}},\ \bibinfo {pages} {064022} (\bibinfo {year} {2006})}\BibitemShut
  {NoStop}%
\bibitem [{\citenamefont {Gutierrez-Pineres}\ and\ \citenamefont
  {Lopez-Monsalvo}(2013)}]{GutierrezPineres:2012yd}%
  \BibitemOpen
  \bibfield  {author} {\bibinfo {author} {\bibfnamefont {Antonio~C.}\
  \bibnamefont {Gutierrez-Pineres}}\ and\ \bibinfo {author} {\bibfnamefont
  {Cesar~S.}\ \bibnamefont {Lopez-Monsalvo}},\ }\bibfield  {title} {\enquote
  {\bibinfo {title} {{A static axisymmetric exact solution of
  $f(R)$-gravity}},}\ }\href {\doibase 10.1016/j.physletb.2012.12.014}
  {\bibfield  {journal} {\bibinfo  {journal} {Phys. Lett.}\ }\textbf {\bibinfo
  {volume} {B718}},\ \bibinfo {pages} {1493--1499} (\bibinfo {year} {2013})},\
  \Eprint {http://arxiv.org/abs/1211.2285} {arXiv:1211.2285 [gr-qc]}
  \BibitemShut {NoStop}%
\bibitem [{\citenamefont {Amirabi}\ \emph {et~al.}(2016)\citenamefont
  {Amirabi}, \citenamefont {Halilsoy},\ and\ \citenamefont
  {Habib~Mazharimousavi}}]{Amirabi:2015aya}%
  \BibitemOpen
  \bibfield  {author} {\bibinfo {author} {\bibfnamefont {Z.}~\bibnamefont
  {Amirabi}}, \bibinfo {author} {\bibfnamefont {M.}~\bibnamefont {Halilsoy}}, \
  and\ \bibinfo {author} {\bibfnamefont {S.}~\bibnamefont
  {Habib~Mazharimousavi}},\ }\bibfield  {title} {\enquote {\bibinfo {title}
  {{Generation of spherically symmetric metrics in f(R) gravity}},}\ }\href
  {\doibase 10.1140/epjc/s10052-016-4164-z} {\bibfield  {journal} {\bibinfo
  {journal} {Eur. Phys. J.}\ }\textbf {\bibinfo {volume} {C76}},\ \bibinfo
  {pages} {338} (\bibinfo {year} {2016})},\ \Eprint
  {http://arxiv.org/abs/1509.06967} {arXiv:1509.06967 [gr-qc]} \BibitemShut
  {NoStop}%
\bibitem [{\citenamefont {Goswami}\ \emph {et~al.}(2014)\citenamefont
  {Goswami}, \citenamefont {Nzioki}, \citenamefont {Maharaj},\ and\
  \citenamefont {Ghosh}}]{PhysRevD.90.084011}%
  \BibitemOpen
  \bibfield  {author} {\bibinfo {author} {\bibfnamefont {Rituparno}\
  \bibnamefont {Goswami}}, \bibinfo {author} {\bibfnamefont {Anne~Marie}\
  \bibnamefont {Nzioki}}, \bibinfo {author} {\bibfnamefont {Sunil.~D.}\
  \bibnamefont {Maharaj}}, \ and\ \bibinfo {author} {\bibfnamefont
  {Sushant~G.}\ \bibnamefont {Ghosh}},\ }\bibfield  {title} {\enquote {\bibinfo
  {title} {Collapsing spherical stars in $f(r)$ gravity},}\ }\href {\doibase
  10.1103/PhysRevD.90.084011} {\bibfield  {journal} {\bibinfo  {journal} {Phys.
  Rev. D}\ }\textbf {\bibinfo {volume} {90}},\ \bibinfo {pages} {084011}
  (\bibinfo {year} {2014})}\BibitemShut {NoStop}%
\bibitem [{\citenamefont {Hendi}\ \emph {et~al.}(2012)\citenamefont {Hendi},
  \citenamefont {Eslam~Panah},\ and\ \citenamefont {Mousavi}}]{Hendi:2011eg}%
  \BibitemOpen
  \bibfield  {author} {\bibinfo {author} {\bibfnamefont {S.~H.}\ \bibnamefont
  {Hendi}}, \bibinfo {author} {\bibfnamefont {B.}~\bibnamefont {Eslam~Panah}},
  \ and\ \bibinfo {author} {\bibfnamefont {S.~M.}\ \bibnamefont {Mousavi}},\
  }\bibfield  {title} {\enquote {\bibinfo {title} {{Some exact solutions of
  F(R) gravity with charged (a)dS black hole interpretation}},}\ }\href
  {\doibase 10.1007/s10714-011-1307-2} {\bibfield  {journal} {\bibinfo
  {journal} {Gen. Rel. Grav.}\ }\textbf {\bibinfo {volume} {44}},\ \bibinfo
  {pages} {835--853} (\bibinfo {year} {2012})},\ \Eprint
  {http://arxiv.org/abs/1102.0089} {arXiv:1102.0089 [hep-th]} \BibitemShut
  {NoStop}%
\bibitem [{\citenamefont {Nashed}(2006)}]{Nashed:2005kn}%
  \BibitemOpen
  \bibfield  {author} {\bibinfo {author} {\bibfnamefont {Gamal G.~L.}\
  \bibnamefont {Nashed}},\ }\bibfield  {title} {\enquote {\bibinfo {title}
  {{Charged axially symmetric solution, energy and angular momentum in tetrad
  theory of gravitation}},}\ }\href {\doibase 10.1142/S0217751X06031478}
  {\bibfield  {journal} {\bibinfo  {journal} {Int. J. Mod. Phys.}\ ,\ \bibinfo
  {pages} {3181--3197}} (\bibinfo {year} {2006})},\ \Eprint
  {http://arxiv.org/abs/gr-qc/0501002} {arXiv:gr-qc/0501002 [gr-qc]}
  \BibitemShut {NoStop}%
\bibitem [{\citenamefont {Bergliaffa}\ and\ \citenamefont
  {Nunes}(2011)}]{PhysRevD.84.084006}%
  \BibitemOpen
  \bibfield  {author} {\bibinfo {author} {\bibfnamefont {Santiago
  Esteban~Perez}\ \bibnamefont {Bergliaffa}}\ and\ \bibinfo {author}
  {\bibfnamefont {Yves Eduardo Chifarelli de~Oliveira}\ \bibnamefont {Nunes}},\
  }\bibfield  {title} {\enquote {\bibinfo {title} {Static and spherically
  symmetric black holes in $f(r)$ theories},}\ }\href {\doibase
  10.1103/PhysRevD.84.084006} {\bibfield  {journal} {\bibinfo  {journal} {Phys.
  Rev. D}\ }\textbf {\bibinfo {volume} {84}},\ \bibinfo {pages} {084006}
  (\bibinfo {year} {2011})}\BibitemShut {NoStop}%
\bibitem [{\citenamefont {Cembranos}\ \emph {et~al.}(2014)\citenamefont
  {Cembranos}, \citenamefont {de~la Cruz-Dombriz},\ and\ \citenamefont
  {Jimeno~Romero}}]{Cembranos:2011sr}%
  \BibitemOpen
  \bibfield  {author} {\bibinfo {author} {\bibfnamefont {J.~A.~R.}\
  \bibnamefont {Cembranos}}, \bibinfo {author} {\bibfnamefont {A.}~\bibnamefont
  {de~la Cruz-Dombriz}}, \ and\ \bibinfo {author} {\bibfnamefont
  {P.}~\bibnamefont {Jimeno~Romero}},\ }\bibfield  {title} {\enquote {\bibinfo
  {title} {{Kerr-Newman black holes in $f(R)$ theories}},}\ }\href {\doibase
  10.1142/S0219887814500017} {\bibfield  {journal} {\bibinfo  {journal} {Int.
  J. Geom. Meth. Mod. Phys.}\ }\textbf {\bibinfo {volume} {11}},\ \bibinfo
  {pages} {1450001} (\bibinfo {year} {2014})},\ \Eprint
  {http://arxiv.org/abs/1109.4519} {arXiv:1109.4519 [gr-qc]} \BibitemShut
  {NoStop}%
\bibitem [{\citenamefont {Aparicio~Resco}\ \emph {et~al.}(2016)\citenamefont
  {Aparicio~Resco}, \citenamefont {de~la Cruz-Dombriz}, \citenamefont
  {Llanes~Estrada},\ and\ \citenamefont {Zapatero~Castrillo}}]{Resco:2016upv}%
  \BibitemOpen
  \bibfield  {author} {\bibinfo {author} {\bibfnamefont {Miguel}\ \bibnamefont
  {Aparicio~Resco}}, \bibinfo {author} {\bibfnamefont {Álvaro}\ \bibnamefont
  {de~la Cruz-Dombriz}}, \bibinfo {author} {\bibfnamefont {Felipe~J.}\
  \bibnamefont {Llanes~Estrada}}, \ and\ \bibinfo {author} {\bibfnamefont
  {Víctor}\ \bibnamefont {Zapatero~Castrillo}},\ }\bibfield  {title} {\enquote
  {\bibinfo {title} {{On neutron stars in $f(R)$ theories: Small radii, large
  masses and large energy emitted in a merger}},}\ }\href {\doibase
  10.1016/j.dark.2016.07.001} {\bibfield  {journal} {\bibinfo  {journal} {Phys.
  Dark Univ.}\ }\textbf {\bibinfo {volume} {13}},\ \bibinfo {pages} {147--161}
  (\bibinfo {year} {2016})},\ \Eprint {http://arxiv.org/abs/1602.03880}
  {arXiv:1602.03880 [gr-qc]} \BibitemShut {NoStop}%
\bibitem [{\citenamefont {Nashed}(2007{\natexlab{a}})}]{Nashed:2006yw}%
  \BibitemOpen
  \bibfield  {author} {\bibinfo {author} {\bibfnamefont {Gamal G.~L.}\
  \bibnamefont {Nashed}},\ }\bibfield  {title} {\enquote {\bibinfo {title}
  {{Kerr-Newman Solution and Energy in Teleparallel Equivalent of Einstein
  Theory}},}\ }\href {\doibase 10.1142/S021773230702141X} {\bibfield  {journal}
  {\bibinfo  {journal} {Mod. Phys. Lett.}\ }\textbf {\bibinfo {volume} {A22}},\
  \bibinfo {pages} {1047--1056} (\bibinfo {year} {2007}{\natexlab{a}})},\
  \Eprint {http://arxiv.org/abs/gr-qc/0609096} {arXiv:gr-qc/0609096 [gr-qc]}
  \BibitemShut {NoStop}%
\bibitem [{\citenamefont {Lobo}\ and\ \citenamefont
  {Oliveira}(2009)}]{PhysRevD.80.104012}%
  \BibitemOpen
  \bibfield  {author} {\bibinfo {author} {\bibfnamefont {Francisco S.~N.}\
  \bibnamefont {Lobo}}\ and\ \bibinfo {author} {\bibfnamefont {Miguel~A.}\
  \bibnamefont {Oliveira}},\ }\bibfield  {title} {\enquote {\bibinfo {title}
  {Wormhole geometries in $f(r)$ modified theories of gravity},}\ }\href
  {\doibase 10.1103/PhysRevD.80.104012} {\bibfield  {journal} {\bibinfo
  {journal} {Phys. Rev. D}\ }\textbf {\bibinfo {volume} {80}},\ \bibinfo
  {pages} {104012} (\bibinfo {year} {2009})}\BibitemShut {NoStop}%
\bibitem [{\citenamefont {Azadi}\ \emph {et~al.}(2008)\citenamefont {Azadi},
  \citenamefont {Momeni},\ and\ \citenamefont {Nouri-Zonoz}}]{Azadi:2008qu}%
  \BibitemOpen
  \bibfield  {author} {\bibinfo {author} {\bibfnamefont {A.}~\bibnamefont
  {Azadi}}, \bibinfo {author} {\bibfnamefont {D.}~\bibnamefont {Momeni}}, \
  and\ \bibinfo {author} {\bibfnamefont {M.}~\bibnamefont {Nouri-Zonoz}},\
  }\bibfield  {title} {\enquote {\bibinfo {title} {{Cylindrical solutions in
  metric f(R) gravity}},}\ }\href {\doibase 10.1016/j.physletb.2008.10.054}
  {\bibfield  {journal} {\bibinfo  {journal} {Phys. Lett.}\ }\textbf {\bibinfo
  {volume} {B670}},\ \bibinfo {pages} {210--214} (\bibinfo {year} {2008})},\
  \Eprint {http://arxiv.org/abs/0810.4673} {arXiv:0810.4673 [gr-qc]}
  \BibitemShut {NoStop}%
\bibitem [{\citenamefont {Nashed}(2010{\natexlab{a}})}]{Nashed:2009hn}%
  \BibitemOpen
  \bibfield  {author} {\bibinfo {author} {\bibfnamefont {Gamal G.~L.}\
  \bibnamefont {Nashed}},\ }\bibfield  {title} {\enquote {\bibinfo {title}
  {{Brane World black holes in Teleparallel Theory Equivalent to General
  Relativity and their Killing vectors, Energy, Momentum and
  Angular-Momentum}},}\ }\href {\doibase 10.1088/1674-1056/19/2/020401}
  {\bibfield  {journal} {\bibinfo  {journal} {Chin. Phys.}\ ,\ \bibinfo {pages}
  {020401}} (\bibinfo {year} {2010}{\natexlab{a}})},\ \Eprint
  {http://arxiv.org/abs/0910.5124} {arXiv:0910.5124 [gr-qc]} \BibitemShut
  {NoStop}%
\bibitem [{\citenamefont {Nashed}(2008)}]{Nashed:2008ys}%
  \BibitemOpen
  \bibfield  {author} {\bibinfo {author} {\bibfnamefont {Gamal Gergess~Lamee}\
  \bibnamefont {Nashed}},\ }\bibfield  {title} {\enquote {\bibinfo {title}
  {{Charged Dilaton, Energy, Momentum and Angular-Momentum in Teleparallel
  Theory Equivalent to General Relativity}},}\ }\href {\doibase
  10.1140/epjc/s10052-007-0511-4} {\bibfield  {journal} {\bibinfo  {journal}
  {Eur. Phys. J.}\ }\textbf {\bibinfo {volume} {C54}},\ \bibinfo {pages}
  {291--302} (\bibinfo {year} {2008})},\ \Eprint
  {http://arxiv.org/abs/0804.3285} {arXiv:0804.3285 [gr-qc]} \BibitemShut
  {NoStop}%
\bibitem [{\citenamefont {Capozziello}\ \emph {et~al.}(2008)\citenamefont
  {Capozziello}, \citenamefont {Stabile},\ and\ \citenamefont
  {Troisi}}]{Capozziello:2007id}%
  \BibitemOpen
  \bibfield  {author} {\bibinfo {author} {\bibfnamefont {S.}~\bibnamefont
  {Capozziello}}, \bibinfo {author} {\bibfnamefont {A.}~\bibnamefont
  {Stabile}}, \ and\ \bibinfo {author} {\bibfnamefont {A.}~\bibnamefont
  {Troisi}},\ }\bibfield  {title} {\enquote {\bibinfo {title} {{Spherical
  symmetry in f(R)-gravity}},}\ }\href {\doibase 10.1088/0264-9381/25/8/085004}
  {\bibfield  {journal} {\bibinfo  {journal} {Class. Quant. Grav.}\ }\textbf
  {\bibinfo {volume} {25}},\ \bibinfo {pages} {085004} (\bibinfo {year}
  {2008})},\ \Eprint {http://arxiv.org/abs/0709.0891} {arXiv:0709.0891 [gr-qc]}
  \BibitemShut {NoStop}%
\bibitem [{\citenamefont {Nashed}(2007{\natexlab{b}})}]{Nashed:2007cu}%
  \BibitemOpen
  \bibfield  {author} {\bibinfo {author} {\bibfnamefont {Gamal Gergess~Lamee}\
  \bibnamefont {Nashed}},\ }\bibfield  {title} {\enquote {\bibinfo {title}
  {{Charged axially symmetric solution and energy in teleparallel theory
  equivalent to general relativity}},}\ }\href {\doibase
  10.1140/epjc/s10052-006-0154-x} {\bibfield  {journal} {\bibinfo  {journal}
  {Eur. Phys. J.}\ ,\ \bibinfo {pages} {851--857}} (\bibinfo {year}
  {2007}{\natexlab{b}})},\ \Eprint {http://arxiv.org/abs/0706.0260}
  {arXiv:0706.0260 [gr-qc]} \BibitemShut {NoStop}%
\bibitem [{\citenamefont {Nashed}(2015)}]{Nashed:2015pda}%
  \BibitemOpen
  \bibfield  {author} {\bibinfo {author} {\bibfnamefont {G.~L.}\ \bibnamefont
  {Nashed}},\ }\bibfield  {title} {\enquote {\bibinfo {title} {{FRW in
  quadratic form of $f(T)$ gravitational theories}},}\ }\href {\doibase
  10.1007/s10714-015-1917-1} {\bibfield  {journal} {\bibinfo  {journal} {Gen.
  Rel. Grav.}\ }\textbf {\bibinfo {volume} {47}},\ \bibinfo {pages} {75}
  (\bibinfo {year} {2015})},\ \Eprint {http://arxiv.org/abs/1506.08695}
  {arXiv:1506.08695 [gr-qc]} \BibitemShut {NoStop}%
\bibitem [{\citenamefont {Cognola}\ \emph {et~al.}(2015)\citenamefont
  {Cognola}, \citenamefont {Rinaldi}, \citenamefont {Vanzo},\ and\
  \citenamefont {Zerbini}}]{PhysRevD.91.104004}%
  \BibitemOpen
  \bibfield  {author} {\bibinfo {author} {\bibfnamefont {Guido}\ \bibnamefont
  {Cognola}}, \bibinfo {author} {\bibfnamefont {Massimiliano}\ \bibnamefont
  {Rinaldi}}, \bibinfo {author} {\bibfnamefont {Luciano}\ \bibnamefont
  {Vanzo}}, \ and\ \bibinfo {author} {\bibfnamefont {Sergio}\ \bibnamefont
  {Zerbini}},\ }\bibfield  {title} {\enquote {\bibinfo {title} {Thermodynamics
  of topological black holes in ${R}^{2}$ gravity},}\ }\href {\doibase
  10.1103/PhysRevD.91.104004} {\bibfield  {journal} {\bibinfo  {journal} {Phys.
  Rev. D}\ }\textbf {\bibinfo {volume} {91}},\ \bibinfo {pages} {104004}
  (\bibinfo {year} {2015})}\BibitemShut {NoStop}%
\bibitem [{\citenamefont {Hendi}\ \emph
  {et~al.}(2014{\natexlab{a}})\citenamefont {Hendi}, \citenamefont
  {Eslam~Panah},\ and\ \citenamefont {Corda}}]{Hendi:2013zba}%
  \BibitemOpen
  \bibfield  {author} {\bibinfo {author} {\bibfnamefont {S.~H.}\ \bibnamefont
  {Hendi}}, \bibinfo {author} {\bibfnamefont {B.}~\bibnamefont {Eslam~Panah}},
  \ and\ \bibinfo {author} {\bibfnamefont {C.}~\bibnamefont {Corda}},\
  }\bibfield  {title} {\enquote {\bibinfo {title} {{Asymptotically Lifshitz
  black hole solutions in F(R) gravity}},}\ }\href {\doibase
  10.1139/cjp-2013-0357} {\bibfield  {journal} {\bibinfo  {journal} {Can. J.
  Phys.}\ }\textbf {\bibinfo {volume} {92}},\ \bibinfo {pages} {76--81}
  (\bibinfo {year} {2014}{\natexlab{a}})},\ \Eprint
  {http://arxiv.org/abs/1309.2135} {arXiv:1309.2135 [gr-qc]} \BibitemShut
  {NoStop}%
\bibitem [{\citenamefont {Hendi}\ \emph
  {et~al.}(2014{\natexlab{b}})\citenamefont {Hendi}, \citenamefont
  {Eslam~Panah},\ and\ \citenamefont {Saffari}}]{Hendi:2014mba}%
  \BibitemOpen
  \bibfield  {author} {\bibinfo {author} {\bibfnamefont {S.~H.}\ \bibnamefont
  {Hendi}}, \bibinfo {author} {\bibfnamefont {B.}~\bibnamefont {Eslam~Panah}},
  \ and\ \bibinfo {author} {\bibfnamefont {R.}~\bibnamefont {Saffari}},\
  }\bibfield  {title} {\enquote {\bibinfo {title} {{Exact solutions of
  three-dimensional black holes: Einstein gravity versus $F(R)$ gravity}},}\
  }\href {\doibase 10.1142/S0218271814500886} {\bibfield  {journal} {\bibinfo
  {journal} {Int. J. Mod. Phys.}\ }\textbf {\bibinfo {volume} {D23}},\ \bibinfo
  {pages} {1450088} (\bibinfo {year} {2014}{\natexlab{b}})},\ \Eprint
  {http://arxiv.org/abs/1408.5570} {arXiv:1408.5570 [hep-th]} \BibitemShut
  {NoStop}%
\bibitem [{\citenamefont {Nashed}(2013{\natexlab{a}})}]{Nashed:uja}%
  \BibitemOpen
  \bibfield  {author} {\bibinfo {author} {\bibfnamefont {Gamal G.~L.}\
  \bibnamefont {Nashed}},\ }\bibfield  {title} {\enquote {\bibinfo {title} {{A
  special exact spherically symmetric solution in f(T) gravity theories}},}\
  }\href {\doibase 10.1007/s10714-013-1566-1} {\bibfield  {journal} {\bibinfo
  {journal} {Gen. Rel. Grav.}\ }\textbf {\bibinfo {volume} {45}},\ \bibinfo
  {pages} {1887--1899} (\bibinfo {year} {2013}{\natexlab{a}})},\ \Eprint
  {http://arxiv.org/abs/1502.05219} {arXiv:1502.05219 [gr-qc]} \BibitemShut
  {NoStop}%
\bibitem [{\citenamefont {Nashed}(2010{\natexlab{b}})}]{Nashed:2015qza}%
  \BibitemOpen
  \bibfield  {author} {\bibinfo {author} {\bibfnamefont {G.~G.~L.}\
  \bibnamefont {Nashed}},\ }\bibfield  {title} {\enquote {\bibinfo {title}
  {{Stationary Axisymmetric Solutions and their Energy Contents in Teleparallel
  Equivalent of Einstein Theory}},}\ }\href {\doibase
  10.1007/s10509-010-0375-1} {\bibfield  {journal} {\bibinfo  {journal}
  {Astrophys. Space Sci.}\ }\textbf {\bibinfo {volume} {330}},\ \bibinfo
  {pages} {173} (\bibinfo {year} {2010}{\natexlab{b}})},\ \Eprint
  {http://arxiv.org/abs/1503.01379} {arXiv:1503.01379 [gr-qc]} \BibitemShut
  {NoStop}%
\bibitem [{\citenamefont {Babichev}\ and\ \citenamefont
  {Langlois}(2010)}]{PhysRevD.81.124051}%
  \BibitemOpen
  \bibfield  {author} {\bibinfo {author} {\bibfnamefont {Eugeny}\ \bibnamefont
  {Babichev}}\ and\ \bibinfo {author} {\bibfnamefont {David}\ \bibnamefont
  {Langlois}},\ }\bibfield  {title} {\enquote {\bibinfo {title} {Relativistic
  stars in $f(r)$ and scalar-tensor theories},}\ }\href {\doibase
  10.1103/PhysRevD.81.124051} {\bibfield  {journal} {\bibinfo  {journal} {Phys.
  Rev. D}\ }\textbf {\bibinfo {volume} {81}},\ \bibinfo {pages} {124051}
  (\bibinfo {year} {2010})}\BibitemShut {NoStop}%
\bibitem [{\citenamefont {Hendi}\ and\ \citenamefont
  {Momeni}(2011)}]{Hendi:2012nj}%
  \BibitemOpen
  \bibfield  {author} {\bibinfo {author} {\bibfnamefont {S.~H.}\ \bibnamefont
  {Hendi}}\ and\ \bibinfo {author} {\bibfnamefont {D.}~\bibnamefont {Momeni}},\
  }\bibfield  {title} {\enquote {\bibinfo {title} {{Black hole solutions in
  F(R) gravity with conformal anomaly}},}\ }\href {\doibase
  10.1140/epjc/s10052-011-1823-y} {\bibfield  {journal} {\bibinfo  {journal}
  {Eur. Phys. J.}\ }\textbf {\bibinfo {volume} {C71}},\ \bibinfo {pages} {1823}
  (\bibinfo {year} {2011})},\ \Eprint {http://arxiv.org/abs/1201.0061}
  {arXiv:1201.0061 [gr-qc]} \BibitemShut {NoStop}%
\bibitem [{\citenamefont {Myrzakulov}\ \emph {et~al.}(2016)\citenamefont
  {Myrzakulov}, \citenamefont {Sebastiani}, \citenamefont {Vagnozzi},\ and\
  \citenamefont {Zerbini}}]{Myrzakulov:2015kda}%
  \BibitemOpen
  \bibfield  {author} {\bibinfo {author} {\bibfnamefont {Ratbay}\ \bibnamefont
  {Myrzakulov}}, \bibinfo {author} {\bibfnamefont {Lorenzo}\ \bibnamefont
  {Sebastiani}}, \bibinfo {author} {\bibfnamefont {Sunny}\ \bibnamefont
  {Vagnozzi}}, \ and\ \bibinfo {author} {\bibfnamefont {Sergio}\ \bibnamefont
  {Zerbini}},\ }\bibfield  {title} {\enquote {\bibinfo {title} {{Static
  spherically symmetric solutions in mimetic gravity: rotation curves and
  wormholes}},}\ }\href {\doibase 10.1088/0264-9381/33/12/125005} {\bibfield
  {journal} {\bibinfo  {journal} {Class. Quant. Grav.}\ }\textbf {\bibinfo
  {volume} {33}},\ \bibinfo {pages} {125005} (\bibinfo {year} {2016})},\
  \Eprint {http://arxiv.org/abs/1510.02284} {arXiv:1510.02284 [gr-qc]}
  \BibitemShut {NoStop}%
\bibitem [{\citenamefont {L\"u}\ \emph
  {et~al.}(2015{\natexlab{a}})\citenamefont {L\"u}, \citenamefont {Perkins},
  \citenamefont {Pope},\ and\ \citenamefont {Stelle}}]{PhysRevD.92.124019}%
  \BibitemOpen
  \bibfield  {author} {\bibinfo {author} {\bibfnamefont {H.}~\bibnamefont
  {L\"u}}, \bibinfo {author} {\bibfnamefont {A.}~\bibnamefont {Perkins}},
  \bibinfo {author} {\bibfnamefont {C.~N.}\ \bibnamefont {Pope}}, \ and\
  \bibinfo {author} {\bibfnamefont {K.~S.}\ \bibnamefont {Stelle}},\ }\bibfield
   {title} {\enquote {\bibinfo {title} {Spherically symmetric solutions in
  higher-derivative gravity},}\ }\href {\doibase 10.1103/PhysRevD.92.124019}
  {\bibfield  {journal} {\bibinfo  {journal} {Phys. Rev. D}\ }\textbf {\bibinfo
  {volume} {92}},\ \bibinfo {pages} {124019} (\bibinfo {year}
  {2015}{\natexlab{a}})}\BibitemShut {NoStop}%
\bibitem [{\citenamefont {L\"u}\ \emph
  {et~al.}(2015{\natexlab{b}})\citenamefont {L\"u}, \citenamefont {Perkins},
  \citenamefont {Pope},\ and\ \citenamefont {Stelle}}]{PhysRevLett.114.171601}%
  \BibitemOpen
  \bibfield  {author} {\bibinfo {author} {\bibfnamefont {H.}~\bibnamefont
  {L\"u}}, \bibinfo {author} {\bibfnamefont {A.}~\bibnamefont {Perkins}},
  \bibinfo {author} {\bibfnamefont {C.~N.}\ \bibnamefont {Pope}}, \ and\
  \bibinfo {author} {\bibfnamefont {K.~S.}\ \bibnamefont {Stelle}},\ }\bibfield
   {title} {\enquote {\bibinfo {title} {Black holes in higher derivative
  gravity},}\ }\href {\doibase 10.1103/PhysRevLett.114.171601} {\bibfield
  {journal} {\bibinfo  {journal} {Phys. Rev. Lett.}\ }\textbf {\bibinfo
  {volume} {114}},\ \bibinfo {pages} {171601} (\bibinfo {year}
  {2015}{\natexlab{b}})}\BibitemShut {NoStop}%
\bibitem [{\citenamefont {Hassa\"{\i}ne}\ and\ \citenamefont
  {Mart\'{\i}nez}(2007)}]{PhysRevD.75.027502}%
  \BibitemOpen
  \bibfield  {author} {\bibinfo {author} {\bibfnamefont {Mokhtar}\ \bibnamefont
  {Hassa\"{\i}ne}}\ and\ \bibinfo {author} {\bibfnamefont {Cristi\'an}\
  \bibnamefont {Mart\'{\i}nez}},\ }\bibfield  {title} {\enquote {\bibinfo
  {title} {Higher-dimensional black holes with a conformally invariant maxwell
  source},}\ }\href {\doibase 10.1103/PhysRevD.75.027502} {\bibfield  {journal}
  {\bibinfo  {journal} {Phys. Rev. D}\ }\textbf {\bibinfo {volume} {75}},\
  \bibinfo {pages} {027502} (\bibinfo {year} {2007})}\BibitemShut {NoStop}%
\bibitem [{\citenamefont {{Misner}}(1963)}]{1963JMP.....4..924M}%
  \BibitemOpen
  \bibfield  {author} {\bibinfo {author} {\bibfnamefont {Charles~W.}\
  \bibnamefont {{Misner}}},\ }\bibfield  {title} {\enquote {\bibinfo {title}
  {{The Flatter Regions of Newman, Unti, and Tamburino's Generalized
  Schwarzschild Space}},}\ }\href {\doibase 10.1063/1.1704019} {\bibfield
  {journal} {\bibinfo  {journal} {Journal of Mathematical Physics}\ }\textbf
  {\bibinfo {volume} {4}},\ \bibinfo {pages} {924--937} (\bibinfo {year}
  {1963})}\BibitemShut {NoStop}%
\bibitem [{\citenamefont {Kagramanova}\ \emph {et~al.}(2010)\citenamefont
  {Kagramanova}, \citenamefont {Kunz}, \citenamefont {Hackmann},\ and\
  \citenamefont {Lämmerzahl}}]{Kagramanova_2010}%
  \BibitemOpen
  \bibfield  {author} {\bibinfo {author} {\bibfnamefont {Valeria}\ \bibnamefont
  {Kagramanova}}, \bibinfo {author} {\bibfnamefont {Jutta}\ \bibnamefont
  {Kunz}}, \bibinfo {author} {\bibfnamefont {Eva}\ \bibnamefont {Hackmann}}, \
  and\ \bibinfo {author} {\bibfnamefont {Claus}\ \bibnamefont {Lämmerzahl}},\
  }\bibfield  {title} {\enquote {\bibinfo {title} {Analytic treatment of
  complete and incomplete geodesics in taub-nut space-times},}\ }\href
  {\doibase 10.1103/physrevd.81.124044} {\bibfield  {journal} {\bibinfo
  {journal} {Physical Review D}\ }\textbf {\bibinfo {volume} {81}} (\bibinfo
  {year} {2010}),\ 10.1103/physrevd.81.124044}\BibitemShut {NoStop}%
\bibitem [{\citenamefont {{Bonnor}}(1969)}]{1969PCPS...66..145B}%
  \BibitemOpen
  \bibfield  {author} {\bibinfo {author} {\bibfnamefont {W.~B.}\ \bibnamefont
  {{Bonnor}}},\ }\bibfield  {title} {\enquote {\bibinfo {title} {{A new
  interpretation of the NUT metric in general relativity}},}\ }\href {\doibase
  10.1017/S0305004100044807} {\bibfield  {journal} {\bibinfo  {journal}
  {Proceedings of the Cambridge Philosophical Society}\ }\textbf {\bibinfo
  {volume} {66}},\ \bibinfo {pages} {145} (\bibinfo {year} {1969})}\BibitemShut
  {NoStop}%
\bibitem [{\citenamefont {Koivisto}\ and\ \citenamefont
  {Kurki-Suonio}(2006)}]{Koivisto:2005yc}%
  \BibitemOpen
  \bibfield  {author} {\bibinfo {author} {\bibfnamefont {Tomi}\ \bibnamefont
  {Koivisto}}\ and\ \bibinfo {author} {\bibfnamefont {Hannu}\ \bibnamefont
  {Kurki-Suonio}},\ }\bibfield  {title} {\enquote {\bibinfo {title}
  {{Cosmological perturbations in the palatini formulation of modified
  gravity}},}\ }\href {\doibase 10.1088/0264-9381/23/7/009} {\bibfield
  {journal} {\bibinfo  {journal} {Class. Quant. Grav.}\ }\textbf {\bibinfo
  {volume} {23}},\ \bibinfo {pages} {2355--2369} (\bibinfo {year} {2006})},\
  \Eprint {http://arxiv.org/abs/astro-ph/0509422} {arXiv:astro-ph/0509422
  [astro-ph]} \BibitemShut {NoStop}%
\bibitem [{\citenamefont {Jaime}\ \emph {et~al.}(2011)\citenamefont {Jaime},
  \citenamefont {Patino},\ and\ \citenamefont {Salgado}}]{Jaime:2010kn}%
  \BibitemOpen
  \bibfield  {author} {\bibinfo {author} {\bibfnamefont {Luisa~G.}\
  \bibnamefont {Jaime}}, \bibinfo {author} {\bibfnamefont {Leonardo}\
  \bibnamefont {Patino}}, \ and\ \bibinfo {author} {\bibfnamefont {Marcelo}\
  \bibnamefont {Salgado}},\ }\bibfield  {title} {\enquote {\bibinfo {title}
  {{Robust approach to f(R) gravity}},}\ }\href {\doibase
  10.1103/PhysRevD.83.024039} {\bibfield  {journal} {\bibinfo  {journal} {Phys.
  Rev.}\ }\textbf {\bibinfo {volume} {D83}},\ \bibinfo {pages} {024039}
  (\bibinfo {year} {2011})},\ \Eprint {http://arxiv.org/abs/1006.5747}
  {arXiv:1006.5747 [gr-qc]} \BibitemShut {NoStop}%
\bibitem [{\citenamefont {Cañate}\ \emph {et~al.}(2016)\citenamefont
  {Cañate}, \citenamefont {Jaime},\ and\ \citenamefont
  {Salgado}}]{Canate:2015dda}%
  \BibitemOpen
  \bibfield  {author} {\bibinfo {author} {\bibfnamefont {Pedro}\ \bibnamefont
  {Cañate}}, \bibinfo {author} {\bibfnamefont {Luisa~G.}\ \bibnamefont
  {Jaime}}, \ and\ \bibinfo {author} {\bibfnamefont {Marcelo}\ \bibnamefont
  {Salgado}},\ }\bibfield  {title} {\enquote {\bibinfo {title} {{Spherically
  symmetric black holes in $f(R)$ gravity: Is geometric scalar hair supported
  ?}}}\ }\href {\doibase 10.1088/0264-9381/33/15/155005} {\bibfield  {journal}
  {\bibinfo  {journal} {Class. Quant. Grav.}\ }\textbf {\bibinfo {volume}
  {33}},\ \bibinfo {pages} {155005} (\bibinfo {year} {2016})},\ \Eprint
  {http://arxiv.org/abs/1509.01664} {arXiv:1509.01664 [gr-qc]} \BibitemShut
  {NoStop}%
\bibitem [{\citenamefont {Nunes}\ \emph {et~al.}(2017)\citenamefont {Nunes},
  \citenamefont {Pan}, \citenamefont {Saridakis},\ and\ \citenamefont
  {Abreu}}]{Nunes:2016drj}%
  \BibitemOpen
  \bibfield  {author} {\bibinfo {author} {\bibfnamefont {Rafael~C.}\
  \bibnamefont {Nunes}}, \bibinfo {author} {\bibfnamefont {Supriya}\
  \bibnamefont {Pan}}, \bibinfo {author} {\bibfnamefont {Emmanuel~N.}\
  \bibnamefont {Saridakis}}, \ and\ \bibinfo {author} {\bibfnamefont {Everton
  M.~C.}\ \bibnamefont {Abreu}},\ }\bibfield  {title} {\enquote {\bibinfo
  {title} {{New observational constraints on $f(R)$ gravity from cosmic
  chronometers}},}\ }\href {\doibase 10.1088/1475-7516/2017/01/005} {\bibfield
  {journal} {\bibinfo  {journal} {JCAP}\ }\textbf {\bibinfo {volume} {1701}},\
  \bibinfo {pages} {005} (\bibinfo {year} {2017})},\ \Eprint
  {http://arxiv.org/abs/1610.07518} {arXiv:1610.07518 [astro-ph.CO]}
  \BibitemShut {NoStop}%
\bibitem [{\citenamefont {Jana}\ and\ \citenamefont
  {Mohanty}(2019)}]{Jana:2018djs}%
  \BibitemOpen
  \bibfield  {author} {\bibinfo {author} {\bibfnamefont {Soumya}\ \bibnamefont
  {Jana}}\ and\ \bibinfo {author} {\bibfnamefont {Subhendra}\ \bibnamefont
  {Mohanty}},\ }\bibfield  {title} {\enquote {\bibinfo {title} {{Constraints on
  $f(R)$ theories of gravity from GW170817}},}\ }\href {\doibase
  10.1103/PhysRevD.99.044056} {\bibfield  {journal} {\bibinfo  {journal} {Phys.
  Rev.}\ }\textbf {\bibinfo {volume} {D99}},\ \bibinfo {pages} {044056}
  (\bibinfo {year} {2019})},\ \Eprint {http://arxiv.org/abs/1807.04060}
  {arXiv:1807.04060 [gr-qc]} \BibitemShut {NoStop}%
\bibitem [{\citenamefont {Mann}\ and\ \citenamefont
  {Stelea}(2006)}]{Mann:2005mb}%
  \BibitemOpen
  \bibfield  {author} {\bibinfo {author} {\bibfnamefont {Robert~B.}\
  \bibnamefont {Mann}}\ and\ \bibinfo {author} {\bibfnamefont {Cristian}\
  \bibnamefont {Stelea}},\ }\bibfield  {title} {\enquote {\bibinfo {title}
  {{New Taub-NUT-Reissner-Nordstrom spaces in higher dimensions}},}\ }\href
  {\doibase 10.1016/j.physletb.2005.10.085} {\bibfield  {journal} {\bibinfo
  {journal} {Phys. Lett.}\ }\textbf {\bibinfo {volume} {B632}},\ \bibinfo
  {pages} {537--542} (\bibinfo {year} {2006})},\ \Eprint
  {http://arxiv.org/abs/hep-th/0508186} {arXiv:hep-th/0508186 [hep-th]}
  \BibitemShut {NoStop}%
\bibitem [{\citenamefont {Clément}\ \emph {et~al.}(2015)\citenamefont
  {Clément}, \citenamefont {Gal'tsov},\ and\ \citenamefont
  {Guenouche}}]{Clement:2015cxa}%
  \BibitemOpen
  \bibfield  {author} {\bibinfo {author} {\bibfnamefont {Gérard}\ \bibnamefont
  {Clément}}, \bibinfo {author} {\bibfnamefont {Dmitri}\ \bibnamefont
  {Gal'tsov}}, \ and\ \bibinfo {author} {\bibfnamefont {Mourad}\ \bibnamefont
  {Guenouche}},\ }\bibfield  {title} {\enquote {\bibinfo {title}
  {{Rehabilitating space-times with NUTs}},}\ }\href {\doibase
  10.1016/j.physletb.2015.09.074} {\bibfield  {journal} {\bibinfo  {journal}
  {Phys. Lett.}\ }\textbf {\bibinfo {volume} {B750}},\ \bibinfo {pages}
  {591--594} (\bibinfo {year} {2015})},\ \Eprint
  {http://arxiv.org/abs/1508.07622} {arXiv:1508.07622 [hep-th]} \BibitemShut
  {NoStop}%
\bibitem [{\citenamefont {Nashed}(2013{\natexlab{b}})}]{Nashed:2013bfa}%
  \BibitemOpen
  \bibfield  {author} {\bibinfo {author} {\bibfnamefont {Gamal G.~L.}\
  \bibnamefont {Nashed}},\ }\bibfield  {title} {\enquote {\bibinfo {title}
  {{Spherically symmetric charged-dS solution in $f(T)$ gravity theories}},}\
  }\href {\doibase 10.1103/PhysRevD.88.104034} {\bibfield  {journal} {\bibinfo
  {journal} {Phys. Rev.}\ }\textbf {\bibinfo {volume} {D88}},\ \bibinfo {pages}
  {104034} (\bibinfo {year} {2013}{\natexlab{b}})},\ \Eprint
  {http://arxiv.org/abs/1311.3131} {arXiv:1311.3131 [gr-qc]} \BibitemShut
  {NoStop}%
\bibitem [{\citenamefont {Hunter}(1999)}]{Hunter:1998qe}%
  \BibitemOpen
  \bibfield  {author} {\bibinfo {author} {\bibfnamefont {C.~J.}\ \bibnamefont
  {Hunter}},\ }\bibfield  {title} {\enquote {\bibinfo {title} {{The Action of
  instantons with nut charge}},}\ }\href {\doibase 10.1103/PhysRevD.59.024009}
  {\bibfield  {journal} {\bibinfo  {journal} {Phys. Rev.}\ ,\ \bibinfo {pages}
  {024009}} (\bibinfo {year} {1999})},\ \Eprint
  {http://arxiv.org/abs/gr-qc/9807010} {arXiv:gr-qc/9807010 [gr-qc]}
  \BibitemShut {NoStop}%
\bibitem [{\citenamefont {Hawking}\ \emph {et~al.}(1999)\citenamefont
  {Hawking}, \citenamefont {Hunter},\ and\ \citenamefont
  {Page}}]{Hawking:1998ct}%
  \BibitemOpen
  \bibfield  {author} {\bibinfo {author} {\bibfnamefont {S.~W.}\ \bibnamefont
  {Hawking}}, \bibinfo {author} {\bibfnamefont {C.~J.}\ \bibnamefont {Hunter}},
  \ and\ \bibinfo {author} {\bibfnamefont {Don~N.}\ \bibnamefont {Page}},\
  }\bibfield  {title} {\enquote {\bibinfo {title} {{Nut charge, anti-de Sitter
  space and entropy}},}\ }\href {\doibase 10.1103/PhysRevD.59.044033}
  {\bibfield  {journal} {\bibinfo  {journal} {Phys. Rev.}\ ,\ \bibinfo {pages}
  {044033}} (\bibinfo {year} {1999})},\ \Eprint
  {http://arxiv.org/abs/hep-th/9809035} {arXiv:hep-th/9809035 [hep-th]}
  \BibitemShut {NoStop}%
\bibitem [{\citenamefont {Bekenstein}(1972)}]{Bekenstein:1972tm}%
  \BibitemOpen
  \bibfield  {author} {\bibinfo {author} {\bibfnamefont {J.~D.}\ \bibnamefont
  {Bekenstein}},\ }\bibfield  {title} {\enquote {\bibinfo {title} {{Black holes
  and the second law}},}\ }\href {\doibase 10.1007/BF02757029} {\bibfield
  {journal} {\bibinfo  {journal} {Lett. Nuovo Cim.}\ }\textbf {\bibinfo
  {volume} {4}},\ \bibinfo {pages} {737--740} (\bibinfo {year}
  {1972})}\BibitemShut {NoStop}%
\bibitem [{\citenamefont {Bekenstein}(1973)}]{Bekenstein:1973ur}%
  \BibitemOpen
  \bibfield  {author} {\bibinfo {author} {\bibfnamefont {Jacob~D.}\
  \bibnamefont {Bekenstein}},\ }\bibfield  {title} {\enquote {\bibinfo {title}
  {{Black holes and entropy}},}\ }\href {\doibase 10.1103/PhysRevD.7.2333}
  {\bibfield  {journal} {\bibinfo  {journal} {Phys. Rev.}\ ,\ \bibinfo {pages}
  {2333--2346}} (\bibinfo {year} {1973})}\BibitemShut {NoStop}%
\bibitem [{\citenamefont {Gibbons}\ and\ \citenamefont
  {Hawking}(1977)}]{Gibbons:1977mu}%
  \BibitemOpen
  \bibfield  {author} {\bibinfo {author} {\bibfnamefont {G.~W.}\ \bibnamefont
  {Gibbons}}\ and\ \bibinfo {author} {\bibfnamefont {S.~W.}\ \bibnamefont
  {Hawking}},\ }\bibfield  {title} {\enquote {\bibinfo {title} {{Cosmological
  Event Horizons, Thermodynamics, and Particle Creation}},}\ }\href {\doibase
  10.1103/PhysRevD.15.2738} {\bibfield  {journal} {\bibinfo  {journal} {Phys.
  Rev.}\ ,\ \bibinfo {pages} {2738--2751}} (\bibinfo {year}
  {1977})}\BibitemShut {NoStop}%
\bibitem [{\citenamefont {Nouicer}(2007)}]{Nouicer:2007pu}%
  \BibitemOpen
  \bibfield  {author} {\bibinfo {author} {\bibfnamefont {Khireddine}\
  \bibnamefont {Nouicer}},\ }\bibfield  {title} {\enquote {\bibinfo {title}
  {{Black holes thermodynamics to all order in the Planck length in extra
  dimensions}},}\ }\href {\doibase 10.1088/0264-9381/24/24/C02,
  10.1088/0264-9381/24/23/014} {\bibfield  {journal} {\bibinfo  {journal}
  {Class. Quant. Grav.}\ }\textbf {\bibinfo {volume} {24}},\ \bibinfo {pages}
  {5917--5934} (\bibinfo {year} {2007})},\ \bibinfo {note} {[Erratum: Class.
  Quant. Grav.24,6435(2007)]},\ \Eprint {http://arxiv.org/abs/0706.2749}
  {arXiv:0706.2749 [gr-qc]} \BibitemShut {NoStop}%
\bibitem [{\citenamefont {Dymnikova}\ and\ \citenamefont
  {Korpusik}(2011)}]{e13121967}%
  \BibitemOpen
  \bibfield  {author} {\bibinfo {author} {\bibfnamefont {Irina}\ \bibnamefont
  {Dymnikova}}\ and\ \bibinfo {author} {\bibfnamefont {Michał}\ \bibnamefont
  {Korpusik}},\ }\bibfield  {title} {\enquote {\bibinfo {title} {Thermodynamics
  of regular cosmological black holes with the de sitter interior},}\ }\href
  {\doibase 10.3390/e13121967} {\bibfield  {journal} {\bibinfo  {journal}
  {Entropy}\ }\textbf {\bibinfo {volume} {13}},\ \bibinfo {pages} {1967--1991}
  (\bibinfo {year} {2011})}\BibitemShut {NoStop}%
\bibitem [{\citenamefont {Fernando}(2017)}]{Fernando:2016ksb}%
  \BibitemOpen
  \bibfield  {author} {\bibinfo {author} {\bibfnamefont {Sharmanthie}\
  \bibnamefont {Fernando}},\ }\bibfield  {title} {\enquote {\bibinfo {title}
  {{Bardeen�de Sitter black holes}},}\ }\href {\doibase
  10.1142/S0218271817500717} {\bibfield  {journal} {\bibinfo  {journal} {Int.
  J. Mod. Phys.}\ ,\ \bibinfo {pages} {1750071}} (\bibinfo {year} {2017})},\
  \Eprint {http://arxiv.org/abs/1611.05337} {arXiv:1611.05337 [gr-qc]}
  \BibitemShut {NoStop}%
\bibitem [{\citenamefont {Katsuragawa}\ and\ \citenamefont
  {Nojiri}(2015)}]{Katsuragawa:2014hda}%
  \BibitemOpen
  \bibfield  {author} {\bibinfo {author} {\bibfnamefont {Taishi}\ \bibnamefont
  {Katsuragawa}}\ and\ \bibinfo {author} {\bibfnamefont {Shin'ichi}\
  \bibnamefont {Nojiri}},\ }\bibfield  {title} {\enquote {\bibinfo {title}
  {{Stability and antievaporation of the Schwarzschild�de Sitter black holes
  in bigravity}},}\ }\href {\doibase 10.1103/PhysRevD.91.084001} {\bibfield
  {journal} {\bibinfo  {journal} {Phys. Rev.}\ ,\ \bibinfo {pages} {084001}}
  (\bibinfo {year} {2015})},\ \Eprint {http://arxiv.org/abs/1411.1610}
  {arXiv:1411.1610 [hep-th]} \BibitemShut {NoStop}%
\bibitem [{\citenamefont {Hawking}(1975)}]{hawking1975}%
  \BibitemOpen
  \bibfield  {author} {\bibinfo {author} {\bibfnamefont {S.~W.}\ \bibnamefont
  {Hawking}},\ }\bibfield  {title} {\enquote {\bibinfo {title} {Particle
  creation by black holes},}\ }\href
  {https://projecteuclid.org:443/euclid.cmp/1103899181} {\bibfield  {journal}
  {\bibinfo  {journal} {Comm. Math. Phys.}\ }\textbf {\bibinfo {volume} {43}},\
  \bibinfo {pages} {199--220} (\bibinfo {year} {1975})}\BibitemShut {NoStop}%
\bibitem [{\citenamefont {Hawking}\ and\ \citenamefont
  {Page}(1983)}]{hawking1983}%
  \BibitemOpen
  \bibfield  {author} {\bibinfo {author} {\bibfnamefont {S.~W.}\ \bibnamefont
  {Hawking}}\ and\ \bibinfo {author} {\bibfnamefont {Don~N.}\ \bibnamefont
  {Page}},\ }\bibfield  {title} {\enquote {\bibinfo {title} {Thermodynamics of
  black holes in anti-de sitter space},}\ }\href {\doibase 10.1007/BF01208266}
  {\bibfield  {journal} {\bibinfo  {journal} {Communications in Mathematical
  Physics}\ }\textbf {\bibinfo {volume} {87}},\ \bibinfo {pages} {577--588}
  (\bibinfo {year} {1983})}\BibitemShut {NoStop}%
\bibitem [{\citenamefont {{Dymnikova}}(1996)}]{Di96}%
  \BibitemOpen
  \bibfield  {author} {\bibinfo {author} {\bibfnamefont {I.~G.}\ \bibnamefont
  {{Dymnikova}}},\ }\bibfield  {title} {\enquote {\bibinfo {title} {{De
  Sitter-Schwarzschild Black Hole:. its Particlelike Core and Thermodynamical
  Properties}},}\ }\href {\doibase 10.1142/S0218271896000333} {\bibfield
  {journal} {\bibinfo  {journal} {International Journal of Modern Physics D}\
  }\textbf {\bibinfo {volume} {5}},\ \bibinfo {pages} {529--540} (\bibinfo
  {year} {1996})}\BibitemShut {NoStop}%
\bibitem [{\citenamefont {Hayward}(2006)}]{Hayward:2005gi}%
  \BibitemOpen
  \bibfield  {author} {\bibinfo {author} {\bibfnamefont {Sean~A.}\ \bibnamefont
  {Hayward}},\ }\bibfield  {title} {\enquote {\bibinfo {title} {{Formation and
  evaporation of regular black holes}},}\ }\href {\doibase
  10.1103/PhysRevLett.96.031103} {\bibfield  {journal} {\bibinfo  {journal}
  {Phys. Rev. Lett.}\ }\textbf {\bibinfo {volume} {96}},\ \bibinfo {pages}
  {031103} (\bibinfo {year} {2006})},\ \Eprint
  {http://arxiv.org/abs/gr-qc/0506126} {arXiv:gr-qc/0506126 [gr-qc]}
  \BibitemShut {NoStop}%
\bibitem [{\citenamefont {Altamirano}\ \emph {et~al.}(2014)\citenamefont
  {Altamirano}, \citenamefont {Kubiznak}, \citenamefont {Mann},\ and\
  \citenamefont {Sherkatghanad}}]{Altamirano:2014tva}%
  \BibitemOpen
  \bibfield  {author} {\bibinfo {author} {\bibfnamefont {Natacha}\ \bibnamefont
  {Altamirano}}, \bibinfo {author} {\bibfnamefont {David}\ \bibnamefont
  {Kubiznak}}, \bibinfo {author} {\bibfnamefont {Robert~B.}\ \bibnamefont
  {Mann}}, \ and\ \bibinfo {author} {\bibfnamefont {Zeinab}\ \bibnamefont
  {Sherkatghanad}},\ }\bibfield  {title} {\enquote {\bibinfo {title}
  {{Thermodynamics of rotating black holes and black rings: phase transitions
  and thermodynamic volume}},}\ }\href {\doibase 10.3390/galaxies2010089}
  {\bibfield  {journal} {\bibinfo  {journal} {Galaxies}\ }\textbf {\bibinfo
  {volume} {2}},\ \bibinfo {pages} {89--159} (\bibinfo {year} {2014})},\
  \Eprint {http://arxiv.org/abs/1401.2586} {arXiv:1401.2586 [hep-th]}
  \BibitemShut {NoStop}%
\bibitem [{\citenamefont {Jawad}\ and\ \citenamefont
  {Shahzad}(2017)}]{Jawad:2017mwt}%
  \BibitemOpen
  \bibfield  {author} {\bibinfo {author} {\bibfnamefont {Abdul}\ \bibnamefont
  {Jawad}}\ and\ \bibinfo {author} {\bibfnamefont {M.~Umair}\ \bibnamefont
  {Shahzad}},\ }\bibfield  {title} {\enquote {\bibinfo {title} {{Effects of
  Thermal Fluctuations on Non-minimal Regular Magnetic Black Hole}},}\ }\href
  {\doibase 10.1140/epjc/s10052-017-4914-6} {\bibfield  {journal} {\bibinfo
  {journal} {Eur. Phys. J.}\ }\textbf {\bibinfo {volume} {C77}},\ \bibinfo
  {pages} {349} (\bibinfo {year} {2017})},\ \Eprint
  {http://arxiv.org/abs/1705.10012} {arXiv:1705.10012 [gr-qc]} \BibitemShut
  {NoStop}%
\bibitem [{\citenamefont {Nashed}(2003)}]{Nashed:2003ee}%
  \BibitemOpen
  \bibfield  {author} {\bibinfo {author} {\bibfnamefont {Gamal G.~L.}\
  \bibnamefont {Nashed}},\ }\bibfield  {title} {\enquote {\bibinfo {title}
  {{Stability of the vacuum nonsingular black hole}},}\ }\href {\doibase
  10.1016/S0960-0779(02)00168-6} {\bibfield  {journal} {\bibinfo  {journal}
  {Chaos Solitons Fractals}\ }\textbf {\bibinfo {volume} {15}},\ \bibinfo
  {pages} {841} (\bibinfo {year} {2003})},\ \Eprint
  {http://arxiv.org/abs/gr-qc/0301008} {arXiv:gr-qc/0301008 [gr-qc]}
  \BibitemShut {NoStop}%
\end{thebibliography}
%

\end{document}